\documentclass[traditabstract]{aa}
\usepackage[varg]{txfonts}
\usepackage{natbib}
\usepackage{graphicx}
\usepackage{psfig}
\usepackage{color}
\bibpunct{(}{)}{;}{a}{}{,}
\definecolor{grey}{rgb}{0.5,0.6,0.7}
\definecolor{green}{rgb}{0.,0.6,0.}

\newcommand{\Fig}[1]{Figure~\ref{fig:#1}}
\newcommand{\be}{\begin{equation}}
\newcommand{\ee}{\end{equation}}

\def\no{\noindent}

\newcommand{\ifm}[1]{\relax\ifmmode#1\else$\mathsurround=0pt #1$\fi}
\newcommand{\kms}{\ifmmode\,{\rm km}\,{\rm s}^{-1}\else km$\,$s$^{-1}$\fi}

\newcommand{\ltsima}{$\; \buildrel < \over \sim \;$}
\newcommand{\lsim}{\lower.5ex\hbox{\ltsima}}
\newcommand{\gtsima}{$\; \buildrel > \over \sim \;$}
\newcommand{\gsim}{\lower.5ex\hbox{\gtsima}}

\def\M11{M_{11}}
\def\V100{V_{100}}
\def\R1{R_{Mpc}}
\def\T6{T_6}

\definecolor{green}{rgb}{0,0.5,0}
\definecolor{brown}{rgb}{0.5,0.3,0}
\definecolor{grey}{rgb}{0.5,0.5,0.5}

\begin{document}

\title{How do galaxies acquire their mass?}
\author{A.~Cattaneo\inst{1,2,3}
\and
G.~A.~Mamon\inst{4,5}
\and
K.~Warnick\inst{3}
\and 
A.~Knebe\inst{3,6}
}
\institute{Laboratoire d'Astrophysique de Marseille (UMR 6110: CNRS \&
  Univ. de Provence), 38, Rue
  Fr\'edr\'eric Joliot-Curie, 13388 Marseille Cedex 13, France
\and
Centre de Recherche Astrophysique de Lyon (UMR 5574: CNRS \& Univ. Claude
Bernard \& ENS Lyon), 9 av. Charles
  Andr{\'e}, 69561 Saint Genis Laval cedex, France
\and
Astrophysikalisches Institut Potsdam, an der Sternwarte 16, 14482 Potsdam,
Germany
\and
Institut d'Astrophysique de Paris (UMR 7095: CNRS \& UPMC), 98 bis
  Boulevard Arago, 75014 Paris, France
\and
Astrophysics \& BIPAC, Department of Physics, University of Oxford, Keble Rd,
Oxford OX1 3RH, UK
\and
Departamento de Fisica Teorica, Modulo C-XI, Universidad Autonoma de Madrid,
28049 Cantoblanco, Madrid, Spain
}
%\email{cattaneo@obs.univ-lyon1.fr}

\date{Received \emph{---} / Accepted \emph{---}}
%%%%%%%%%%%%%%%%%%%%%%%%%%%%%%%%%%%%%%
\abstract{We study the growth of galaxy masses, via gas accretion and galaxy mergers.
We introduce a toy model that
describes (in a single equation) how much baryonic mass is accreted and retained
into galaxies as a function of halo mass and redshift.
In our model, the evolution of the baryons differs from that of the dark
matter because
1) gravitational shock heating and AGN jets suppress gas accretion mainly
above a critical halo mass of $M_{\rm shock}\sim 10^{12}M_\odot$; 2) the
intergalactic medium after reionisation is too hot for accretion onto haloes
with circular velocities $v_{\rm circ} \la 40 {\rm\,km\,s}^{-1}$; 3)
stellar feedback drives gas out of haloes, mainly those with $v_{\rm circ} \la
120{\rm\,km\,s}^{-1}$.  We run our model on the 
merger trees of the haloes and subhaloes of a high-resolution dark matter
cosmological simulation.  The galaxy mass is taken as the maximum between the
mass given by the toy model and the sum of the masses of its progenitors
(reduced by tidal stripping).

Designed to reproduce the present-day stellar mass function of galaxies, our
model matches fairly well the evolution of the cosmic stellar density.
It leads to the same $z$=0 relation between central galaxy stellar and halo
mass as the one
found by abundance matching and also as that previously measured at high mass
on SDSS centrals.
Our model also predicts a bimodal distribution (centrals and satellites) of
stellar masses for given halo mass,
in very good agreement with SDSS observations.

The relative importance of mergers depends strongly on stellar mass (more
than on halo mass).  Massive galaxies with $m_{\rm stars}>m_{\rm crit}\sim
\Omega_{\rm b}/\Omega_{\rm m}M_{\rm shock}\sim 10^{11}M_\odot$ acquire most
of their final mass through mergers (mostly major and gas-poor), as expected from
our model's shutdown of gas accretion at high halo masses.  However, although
our mass resolution should see the effects of mergers down to $m_{\rm
  stars}\simeq 10^{10.6} h^{-1}M_\odot$, we find that mergers are rare for
$m_{\rm stars}\lsim 10^{11}\,h^{-1} M_\odot$. This is a consequence of the
curvature of the stellar vs. halo mass relation set by the physical processes
of our toy model and found with abundance matching.  
So gas accretion must be the dominant growth mechanism for
intermediate and low mass galaxies, including dwarf ellipticals in clusters.  
The contribution of 
galaxy mergers terminating in haloes with mass $M_{\rm halo}<M_{\rm shock}$ (thus presumably
gas-rich) 
to the mass buildup
of galaxies is small at all masses, but accounts for the
bulk of the growth of ellipticals of intermediate mass ($\sim 10^{10.5}
h^{-1} M_\odot$), which we predict must be the typical mass of ULIRGs.
}


\titlerunning{How do galaxies acquire their mass?}
\authorrunning{A. Cattaneo, G.~A. Mamon, K. Warnick, A. Knebe}

\maketitle

\label{firstpage}

%%%%%%%%%%%%%%%%%%%%%%%%%%%%%%%

\section{Introduction}
\label{sec:intro}

In the standard theory \citep{white_rees78,blumenthal_etal84,white_frenk91},
galaxy formation is a two-stage process.  The gravitational instability of
primordial density fluctuations forms dark matter haloes that merge
hierarchically into larger and larger structures.  Galaxies grow within these
haloes (i) by accreting gas that dissipates and falls to the centre of the
gravitational potential wells of the dark matter and (ii) by merging with
other galaxies after their haloes have become part of larger structures.
These two paths explain why there are two galaxy morphological types: in the
standard theory, gas accretion is the mechanism that builds up disc galaxies
(i.e. spirals; \citealp{fall_efstathiou80}), while galaxy merging is the
mechanism by which elliptical galaxies acquire the bulk of their mass
\citep{toomre77,mamon92}.

In the present article, we wish to revisit the question of how galaxies
acquired their present-day stellar mass. Was it mainly through mergers or
through smooth gas accretion?  
Were the galaxies merging together mainly gas-rich (often called `wet' mergers)
or gas-poor (`dry' mergers)?
How
important are \emph{major} mergers involving galaxies of similar mass,
compared to more \emph{minor} ones involving very different mass galaxies?
What are the respective roles of feedback mechanisms such as reionisation of
the IGM \citep{rees86}, supernovae explosions \citep{dekel_silk86}, jets from
active galactic nuclei (AGN, \citealp{silk_rees98}), and the absence or
presence of shock fronts between the infalling and virialised gas, depending
on the mass of the halo \citep{birnboim_dekel03}?  How do the relative
impacts of these physical processes depend on the final (observable) stellar
mass of the galaxy?

To answer these questions, one must first estimate the build-up of dark matter haloes,
either through Monte-Carlo merger-trees
\citep{lacey_cole93,somerville_kolatt99,neistein_dekel08} or through
cosmological N-body simulations, which also provide spatial information.
If we assume that the position of a galaxy is tracked by the centre of mass of its halo,
then the halo merger rate will also give us the rate at which galaxies merge
(assuming the knowledge of the evolution of haloes once they become subhaloes
of larger ones).
It is more difficult to compute the stellar masses of galaxies prior
to merging,
which determine the contribution from gas accretion,
 because they depend not only on the gas mass that 
accretes onto galaxies but also on feedback processes that eject gas
from galaxies.

There are two approaches to deal with this complexity.  Semi-analytic models
(SAMs, \citealp{kauffmann_etal93,cole_etal94}; also
\citealp{cattaneo_etal06,Bower+06,croton_etal06,somerville_etal08,lofaro_etal09}, 
and references therein) use simple recipes to
describe the various physical processes affecting the baryons within haloes, and
predict many observables. However, even the simplest 
SAMs quickly reach considerable complexity and involve a large number of free
parameters.

The halo occupation distribution (HOD) approach
\citep{berlind_weinberg02,conroy_etal06,yang_etal09} does not make any assumption about the
physics that govern the number \citep{berlind_weinberg02}, the luminosity
\citep{yang_etal03} or the stellar mass \citep{yang_etal09} of galaxies
within haloes.  Instead, in the HOD approach, one computes either the mean
(abundance matching, see \citealp{marinoni_hudson02,conroy_etal06}) or
the statistical
distributions that these properties must have as a function of halo mass and
redshift to reproduce the observational data (assuming that the properties of
dark matter haloes, i.e. the mass function and clustering, are modelled
correctly).  However, HOD models cannot tell us how much of this mass comes
from gas accretion and how much comes from mergers.

Our approach is intermediate between the two.  
We parameterise, in a single equation, the mass of stars formed after gas accretion as a
function of halo mass and redshift, 
$m_{\rm stars} = \widetilde m_{\rm stars}[M_{\rm halo},v_{\rm circ}(M_{\rm halo},z)]$,
where $\widetilde m_{\rm stars}$ takes into account various feedback
effects that quench gas accretion and subsequent star formation.  We compute
galaxy mergers by following dark matter substructures.  When the latter are
no longer resolved, we merge the galaxies on a dynamical friction timescale
using a formula calibrated on cosmological hydrodynamic simulations
\citep{jiang_etal08},
thus allowing the formation of galaxies with $m_{\rm stars} > \widetilde m_{\rm stars}[M_{\rm halo},v_{\rm circ}(M_{\rm halo},z)]$.

Not only is our approach much simpler than fully fledged SAMs 
(we shall see that it contains only four parameters),
it is also simpler than ``lighter" models 
such as the one recently proposed by \cite{neistein_weinmann10}, who
parameterised the dependence of the time derivatives of the masses of the
stellar, cold and hot gas components of galaxies as a function of halo mass
and redshift.  
Admittedly, the simplicity of our approach limits its scope. But we believe
that it makes robust predictions on the mass growth of galaxies, which
provide a stepping stone to more complex analyses.
At the same time, differently from HOD, our model allow us to follow mergers and
thus to separate growth via gas accretion and growth via mergers.

The structure of the article is as follows. In Section~2, we present the
N-body simulation used to follow the gravitational evolution of the
dark matter, and two simple formulae, one that expresses the
timescale for orbital decay of 
unresolved subhaloes, and the other that expresses the stellar mass
at the centre of each halo if galaxies grew only by gas accretion.  
In Section~3, we present our results for the
relative importance of gas accretion and mergers as a function of galaxy
mass.  Section~4 discusses our results and summarises our conclusions.

\section{The galaxy formation model}
\label{sec:toy}

\subsection{Gravitational evolution of the dark matter}

The hierarchical formation and clustering of dark matter haloes is followed
by means of a cosmological $N$-body simulation. The computational volume of
the simulation presented here is a cube with side length $L=50h^{-1}{\rm
  Mpc}$ and periodic boundary conditions. The values of the cosmological
parameters used to generate the initial conditions at redshift $z=50$ by
means of the Zel'dovich approximation \citep{efstathiou_etal85} and to run
the simulation are those from the Wilkinson Microwave Anisotropy Probe's
5th-year analysis (WMAP5) combined with results from type Ia supernovae and
baryonic acoustic oscillations \citep{komatsu_etal09}: $\Omega_{\rm m} =
0.279$, $\Omega_{\rm \Lambda} = 0.721$, $\Omega_{\rm b} = 0.046$, $h = 0.70$,
$\sigma_8 = 0.817$ and $n_s = 0.96$. The simulation was run with the tree-PM
code GADGET2 \citep{Springel05} at a resolution of $512^3$ particles. We
saved the outputs at 101 timesteps equally spaced in time ($\Delta
t=92h^{-1}{\rm\,Myr}$) from $z=10$ to $z=0$. The number of timesteps was
chosen so that $\Delta t$ is smaller than any merging timescale.  The
particle mass is $7.3\times 10^7h^{-1}\,M_\odot$ and the force softening is
$2 \, h^{-1} \, \rm kpc$.

The MPI version of the \texttt{AMIGA}\footnote{Adaptive Mesh Investigations
  of Galaxy Assembly} halo finder, \texttt{AHF}
\citep{knollmann_knebe09},\footnote{AMIGA's Halo Finder} 
was used to identify haloes and sub-haloes in each
of the saved outputs
\footnote{\texttt{AHF} can be downloaded from
  \texttt{http://popia.ft.uam.es/AMIGA}}.  \texttt{AHF} is an improvement of
the \texttt{MHF} halo finder \citep{gill_etal04}, which locates peaks in an
adaptively smoothed density field as prospective halo centres. 
For each of these density peaks, we determined the particle with the lowest gravitational
potential and considered the surrounding particles sorted in distance to
iteratively find the largest collection of bound particles encompassing the density peak,
using their mean velocity as the kinetic reference. Only peaks with at least 20
bound particles are considered as haloes and retained for further analysis.
Therefore, the minimum halo mass is $1.5\times 10^9h^{-1}\,M_\odot$
(corresponding to $v_{\rm circ}=15 \, \rm km \, s^{-1}$).  Our halo finding
algorithm automatically identifies haloes, sub-haloes, sub-sub-haloes, and so
on (see \citealp{knollmann_knebe09} for the details of the algorithm).

For each halo at epoch $z$, we compute the virial radius $r_{\rm vir}$, which
is the radius where the mean density drops below $\Delta(z)$ times the
critical density of the Universe at redshift $z$.  The threshold $\Delta(z)$
is computed using the spherical top-hat collapse model and is a function of
both cosmological model and time (\citealp{nakamura96}, cited in
\citealp{kitayama_suto96}; \citealp{gross97,bryan_norman98}).  For the
cosmology that we are using, $\Delta = 98$ at $z=0$, i.e. the overdensity at
the virial radius is 352 times the present-day mean density of the Universe.

After each halo has been identified by \texttt{AHF} in this manner, we define
sub-haloes to be haloes (defined with the same overdensity $\Delta$) 
that lie within the virial region of a more massive halo, the
so-called host halo, of mass $M_{\rm halo}^{\rm host}$.  As sub-haloes are
embedded within their host halo, their own density profile usually shows a
characteristic upturn at the radius $r_t\lsim r_{\rm vir}$ where the density
profile becomes dominated by the surrounding host halo
\footnote{The actual density profile of sub-haloes
 after the removal of the host's background drops faster than for isolated
 haloes \citep[e.g.][]{kazantzidis_etal04}. Only when measured within the
 background still present shall we find the characteristic upturn used here
 to define the truncation radius $r_t$.}.  We use this truncation radius
as the outer edge of the sub-halo. Halo and sub-halo properties
(i.e. mass, density profile, velocity dispersion, rotation curve) are
calculated using the gravitationally bound particles inside either the virial
radius $r_{\rm vir}$ for a host halo or the truncation radius $r_t$ for a
sub-halo.

We build merger trees by cross-correlating haloes in consecutive simulation
outputs.  For this purpose, we use a tool that comes with the \texttt{AHF}
package, called \texttt{MergerTree}. As the name suggests, it serves the
purpose of identifying corresponding objects in a same simulation at
different redshifts. We follow each halo (either host or sub-halo) identified
at redshift $z$=0 backwards in time. The direct progenitor at the previous
redshift is the one that shares the greatest number of particles with the
present halo \textit{and} is closest to it in mass. The latter criterion is
important for sub-haloes as all their particles are also part of the host
halo, but there is normally a large gap between sub-halo and host halo
masses.

In the study presented here, we walk along the tree starting at redshift
$z=10$ and moving toward $z$=0. Sub-haloes are followed within the
environment of their respective hosts until the point where \texttt{AHF} can
no longer resolve them either because they have been tidally disrupted or
because they have merged with their host (of course, many sub-haloes survive
until $z$=0).  Section~\ref{sec:galaxy_mergers} below describes how we follow
sub-haloes when we can no longer resolve them numerically.

\subsection{Growth via galaxy mergers}
\label{sec:galaxy_mergers}

We associate galaxies to the smallest resolved dark matter substructures. 
We say that two haloes have merged at the cosmic time $t_{\rm m}^{\rm haloes}$ when a subhalo
is no longer resolved in the N-body simulation (Fig.~1). 
With this definition, the merging of two dark matter haloes is a necessary
but non-sufficient 
condition for the merging of their central galaxies because generally
a sub-halo ceases to be resolved before reaching the centre of its host halo.
Let $t_{\rm df}$ be the time the sub-halo takes to reach the centre after it
is no longer resolved. 
The galaxy at the centre of the sub-halo has merged with the galaxy at the
centre of the host halo by cosmic time $t$ if
$t_{\rm m}^{\rm haloes}+t_{\rm df}\le t$.
If this condition is satisfied, then the subhalo's central galaxy is merged with
that of the host halo at the first timestep for which $t> t_{\rm m}^{\rm haloes} + t_{\rm
    df}$. Otherwise, the galaxy properties are frozen after $t_{\rm m}^{\rm haloes}$.

\begin{figure}
% GAM SM orbit orbit
\noindent
\centering
\includegraphics[width=0.8\hsize,bb=75 213 585 705]{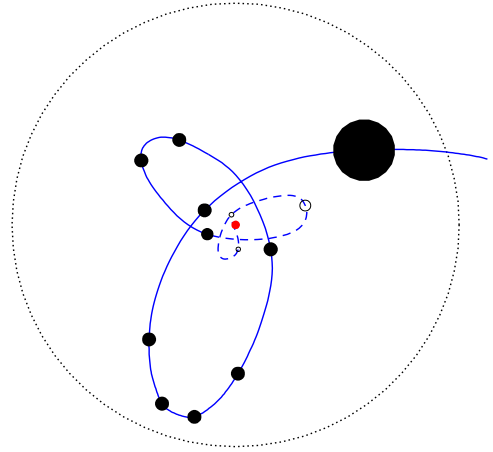}
\caption{Sketch illustrating the merger of two haloes. Once the smaller halo
  penetrates the larger one (whose central galaxy is denoted by a \emph{red
    circle}), 
it becomes a sub-halo. Its orbit (adapted from
  an $N$-body simulation by \cite{CMG99}), shown by the
  \emph{blue curve} (with positions shown for equal timesteps), 
decays by dynamical friction and its size, shown by the
  \emph{circles}, is reduced by tidal stripping from the larger halo's
  potential.  The \emph{solid part of the blue curve} is followed directly by
  the N-body simulation, while the \emph{dashed part} shows the trajectory
  that the sub-halo is expected to take after it  (\emph{open circles}) is 
no longer resolved in the
  simulation.  In our analysis, 
the duration of the motion along the dashed stretch of the blue
  curve is computed analytically from Eqs.(\ref{tdf}) and (\ref{jiang}).}
\label{fig:DynamicalFriction}
\end{figure}

To compute the time for the satellite's orbit to decay to the centre, we use
Chandrasekhar's (\citeyear{Chandrasekhar43}) dynamical friction time, which
can be written as
\begin{equation}
\label{tdf}
t_{\rm df}=A\,{r^2v_{\rm circ}\over{\rm G}M_{\rm sat}\,{\rm ln}\Lambda},
\end{equation}
(e.g. \citealp{binney_tremaine08}), where $r$ is the sub-halo's orbital
radius at the time $t_{\rm m}^{\rm haloes}$ when it ceases to be resolved, $v_{\rm circ}$
is the circular velocity at radius $r$, ${\rm ln}\Lambda$ is the so-called Coulomb
logarithm, and $A$ is a factor of order unity that depends on orbital
eccentricity (\citealp*{CMG99}, and see below).
In the following two paragraphs, we explain our choice for $\ln\Lambda$ and $A$.

\citet{jiang_etal08} used high-resolution cosmological hydrodynamical
simulations to show that, once a halo of mass $M_{\rm sat}$ penetrates the virial sphere of a
larger one, of mass $M_{\rm halo}$ (and thus becomes a sub-halo, i.e. satellite), the
time it takes to fall to the centre of its parent halo is given by
equation~(\ref{tdf}), where $r$ and $v_{\rm circ}$ are measured at the virial
radius $r_{\rm vir}$ of the host halo and where the Coulomb logarithm is
$\ln \Lambda = \ln (1+M_{\rm halo}/M_{\rm sat})$.
We thus assume that for $r<r_{\rm vir}$, the appropriate Coulomb logarithm is the
one computed using 
\begin{equation}
\Lambda = 1+{M_{\rm halo}(r)\over M_{\rm sat}}
\ ,
\label{Coulomb}
\end{equation}
where $M_{\rm halo}(r)$ is the
halo mass enclosed within the sphere of radius $r$.

The normalisation $A$ of the dynamical friction time is often chosen to be
independent of the shape of the satellite's orbit (e.g. 1.17 according to
\citealp{binney_tremaine08}).  But for given orbit apocentre, the time of
orbital decay is expected to be shorter for satellites on radial orbits,
which feel more dynamical friction as they reach the high density region near
the centre of their halo.  \citeauthor{jiang_etal08} calibrated this effect
and found
\begin{equation} 
\label{jiang}
A(\epsilon)=1.17\,(0.94\,\epsilon^{0.6}+0.6) \ ,
\end{equation}
where $\epsilon = J/J_{\rm circ}(E)$ is the orbital circularity, that is, the
ratio of angular momentum to that of a circular orbit of the same energy
($\epsilon=0$ for a radial orbit and $\epsilon=1$ for a circular orbit).
\citeauthor{jiang_etal08}'s parametrisation of the Coulomb logarithm and
normalisation of the orbital decay time, based upon hdyrodynamical
cosmological N-body simulations, implicitly takes into account the
greater concentration of baryons with respect to the dark matter as well as
the effects of tidal stripping on the dynamical friction time.

We now explain how we compute in our model the value of $\epsilon$ that we insert into
Eq.~(\ref{jiang}).
The radial coordinate $r$,
the orbital speed $v$, and the radial and tangential velocities $v_r$ and
$v_t=\sqrt{v^2-v_r^2}$ of the sub-halo with respect to the centre of mass of
the host halo are extracted from the N-body simulation at the last timestep
when the sub-halo is resolved.  We compute the sub-halo's angular momentum
and energy per unit mass. They are, respectively, $J=r\,v_t$ and
$E=1/2\,v^2+\phi(r)$, where $\phi(r)$ is the gravitational potential of the
host halo that we compute assuming an NFW model \citep{navarro_etal96}.  This
requires to know the halo concentration.  The latter could be evaluated
directly from the particle distribution but this may lead to noise,
particularly in low-mass haloes, which contain fewer particles.  Instead, we
use the most recent determination of the concentration-mass relation for
regular haloes in $\Lambda$CDM simulations with WMAP-5th year cosmological
parameters \citep{maccio_etal08}: $c=6.1\,(M_{\rm
  halo}/10^{14}h\,M_\odot)^{-0.094}$. With this assumption we compute $J_{\rm
  circ}(E)$ by solving the equation $E=1/2\,v_{\rm circ}^2+\phi(r)$ for $r$
where the squared circular velocity is $v_{\rm circ}^2={\rm G}\,M_{\rm
  halo}^{\rm host}(r)/r$, and where $M_{\rm halo}^{\rm host}(r)$ is also
computed with the NFW model.  This equation has a solution for $E<0$. When
$E\ge 0$, the sub-structure is not gravitationally bound to the host, so the
satellite and the central galaxy do not merge.

In our analysis, the galaxy merging history is determined fully and solely by
the underlying evolution of the dark matter component. It does not depend on
the efficiency with which gas dissipates, sinks to the centre and makes
stars.  This is why we have discussed galaxy mergers immediately after
presenting the N-body simulation. The processes that determine the growth of
luminous galaxies via accretion within dark matter haloes are considered in
the following subsection.

\subsection{Growth via gas accretion}
\label{sec:accr}
We  compute the stellar mass of a galaxy, $m_{\rm stars}$, as a
function of $M_{\rm halo}$ using a model that
is an improvement over the one introduced by
\citet{cattaneo01}.  Despite its simplicity, that model was shown to be in
reasonable agreement with the galaxy luminosity function and the cosmic
evolution of the star formation rate (SFR) density.  

The key assumption is that, in the 
absence of mergers, the mass of the stars in a dark matter halo
is basically a function of two quantities only: the halo mass $M_{\rm halo}$ and the halo circular velocity $v_{\rm circ}(M_{\rm halo},z)$.

The effects of shock heating and feedback enter our model through the form of this function.
Cosmic reionisation suppresses gas
accretion and star formation in haloes below a minimum circular velocity
$v_{\rm reion}$, stellar feedback mitigates star formation in haloes below a
characteristic circular velocity $v_{\rm SN}$, and gravitational shock
heating coupled to black hole feedback suppresses gas accretion and star
formation above a critical halo mass of $M_{\rm shock}$.  

The stellar mass of a galaxy at any given time is assumed to be given by
\begin{equation}
\label{M1new}
m_{\rm stars} = {\rm max}\left\{  \widetilde m_{\rm stars}[M_{\rm halo},v_{\rm circ}(M_{\rm halo},z)],\,
\sum_{\rm prog} m_{\rm stars}^{\rm (prog)}s_{\rm prog}\right\},
\end{equation}
where $\sum_{\rm prog} m_{\rm stars}^{\rm (prog)}$ is the sum of the stellar
masses of all progenitors and $0<s_{\rm prog}\le 1$ is a factor that
accounts for the tidal stripping of the mass of a satellite progenitor prior
to merging.  For the main progenitor, $s_{\rm 
  prog}=1$. The way in which $s_{\rm prog}$ is computed for the other progenitors is
discussed in Section \ref{tidal_str}.

Our model does not contain any rate and does not integrate any differential
equations. However, Eq.~(\ref{M1new}) implies that the stellar mass at
timestep $i$ can be larger than the value computed from the function
$\widetilde m_{\rm stars}$ at timestep $i$ if the stellar mass at timestep
$i-1$ is larger than the mass computed from the function $\widetilde m_{\rm
  stars}$ at timestep $i$. This has a simple physical reason: besides the
effects of stellar evolution (neglected in our model), the mass of the stars
that have already formed and are already in a galaxy cannot decrease, even if
the potential well becomes shallower.

While equation~(\ref{M1new}) suggests that the final halo mass is independent
of the mass accretion history (MAH) of its halo, our prevention of any decrease in
stellar mass  implies some dependence on the halo MAH: 
A galaxy with a low-mass present-day halo might
have never formed stars if its halo circular velocity was always below the
threshold for galaxy (star) formation (see
 Sect.~\ref{secreion} below). Another galaxy with the same halo mass at
 $z=0$ may have acquired stars if its halo mass had grown fast enough for its
 circular velocity to break the threshold for galaxy formation, and in our
 model it would keep its stellar mass.
The implications for the results of our model will be
discussed in Section~3.

We now describe how we choose the functional form of $\widetilde m_{\rm
  stars}$ (the rest of this 
Section) and $s_{\rm prog}$ (Section~\ref{tidal_str}).

\subsubsection{Shock heating and AGN feedback}
\label{secshock}
The accretion of gas onto low mass haloes is understood to proceed in the form
of cold filaments \citep{keres_etal05} that fall onto the disc, where the new
material is shock heated, but cools very rapidly thanks to the very high gas
density of the disc. However, as shown by \cite{birnboim_dekel03}, above a critical halo
mass $M_{\rm shock} \sim 10^{12} M_\odot$ (roughly independent in time, see
Fig.~2 of \citealp{dekel_birnboim06}), 
the condition for the propagation of a stable shock is satisfied, and
the post-shock gas builds up a hot atmosphere that cools inefficiently
(see also \citealp{oser_etal10}).
Moreover, AGN
couple to the hot gas preventing it from cooling down again
(see the review of \citealp{cattaneo_etal09}, and references
therein, in particular \citealp{croton_etal06} for the first explicit
implementation in a SAM). 
Following \citet{cattaneo01}, we model these two effects by assuming that
the mass  accreted by the galaxy from the start of the simulation to the
timestep under consideration is, to first order,
\begin{equation}
m_{\rm accr}^{(1)}={f_{\rm b}\,M_{\rm halo}\over 1+M_{\rm halo}/M_{\rm
    shock}} \ ,
\label{mshock}
\end{equation}
for $v_{\rm circ}\gg v_{\rm reion}$, where $f_{\rm b}=\Omega_{\rm
  b}/\Omega_{\rm m}$ is the mean baryonic fraction of the Universe.  
In other words, $m_{\rm
  accr}^{(1)}\simeq f_{\rm b}\,M_{\rm halo}$ when $M_{\rm halo}\ll M_{\rm
  shock}$, while $m_{\rm accr}^{(1)}\simeq f_{\rm b}\,M_{\rm shock}$ when
$M_{\rm halo}\gg M_{\rm shock}$.  In the latter case, the galaxy mass grows
very slowly with increasing halo mass (except for the effects of mergers,
which are not considered in the calculation of the accreted
mass).\footnote{Admittedly, the denominator of equation~(\ref{mshock}) is ad
  hoc. We tried powers other than unity and other forms, but the form of
  equation~(\ref{mshock}) was the simplest one that gave us a good match to
  the high-end of the mass function (see Sect.~\ref{sec:bestfit}).}

\subsubsection{Reionisation} 
\label{secreion}
Heating by the photoionising UV background raises the
entropy of the gas and suppresses the concentration of baryons in shallow
potential wells
(\citealp*{ikeuchi86,rees86,blanchard_etal92} --- for Compton heating ---,
\citealp*{efstathiou92,gnedin00,kravtsov_etal04}). 
Here we take a phenomenological approach and model the effects of
photoionisation heating by introducing a lower circular velocity cut-off at
$v_{\rm circ}\sim v_{\rm reion}$ that mimics the shutoff of gas infall when
$T_{\rm vir} < T_{\rm IGM}$ \citep{blanchard_etal92,thoul_weinberg96}.
Cosmological hydrodynamical simulations give conflicting answers on the
temperature of the intergalactic medium (IGM) in the intermediate density
regions outside the virial radius, from which gas should fall in
(\citealp{McQuinn+09}, and references therein).  We assume that this IGM
temperature does not vary with redshift at $z\lsim z_{\rm reion}>10$
($z\simeq 10$ is where our calculations start), which is in rough agreement
with hydrodynamical 
simulations (\citealp{McQuinn+09}, and references therein). We also
suppose that the mass $m_{\rm accr}$ that can flow to the centre in a halo of
mass $M_{\rm halo}$ is suppressed with respect to the first-order accreted
mass by a factor of $1-(v_{\rm reion}/v_{\rm circ})^2$ at $v_{\rm circ}\ge
v_{\rm reion}$, so that:
\begin{equation}
m_{\rm accr} = \left [1-\left ({v_{\rm reion}\over
   v_{\rm circ}}\right)^2\right]\,m_{\rm accr}^{(1)} \ .
\label{mreion}
\end{equation}
At $v_{\rm circ}< v_{\rm reion}$ the suppression is
total ($m_{\rm accr}=0$).  This extreme approximation is
\emph{ad hoc}, but it is not
unreasonable given the high mass-to-light ratios measured in low-mass
objects and is in agreement with hydrodynamical simulations
\citep{thoul_weinberg96}.\footnote{\cite{gnedin00} have used more refined
  hydrodynamical simulations to find that the suppression by reionisation
  scales as $v_{\rm circ}^9$ (see also \citealp{okamoto_etal08}), but this
  should produce negligible differences in the mass range of galaxies studied
  here.}

\subsubsection{Supernovae} 
\label{secsn}

When the circular velocity becomes larger than $v_{\rm
  reion}$, the gas starts flowing to the centre and making stars, which feed
energy back to the interstellar medium mainly via type II supernova
explosions. The energy fed back to the interstellar medium is proportional to
the stellar mass of the galaxy,  $\widetilde m_{\rm stars}$.  
A fraction of this energy is used to drive an
outflow. Let $m_{\rm wind}$ be the total mass ejected from a galaxy from the
start of the simulation to the timestep under consideration.
For a given wind kinetic energy, $1/2\,m_{\rm wind}v_{\rm wind}^2$,
$m_{\rm wind}$ is maximum when the wind speed, $v_{\rm wind}$,
is minimum. However, the wind speed cannot be lower than the escape velocity,
which is proportional to the circular velocity $v_{\rm circ}$, because
otherwise the gas would not flow out. The maximum outflow condition thus implies
that $m_{\rm wind}=(v_{\rm SN}/v_{\rm circ})^2 \widetilde m_{\rm stars}$
\citep{dekel_silk86}, where $v_{\rm SN}^2$ is a proportionality constant, the
value of which is to be determined by fitting the galaxy mass function and is
physically related to the supernova energy that goes into wind kinetic
energy.  We also require mass conservation: $m_{\rm wind}+\widetilde m_{\rm
  stars}=m_{\rm accr}$. In other words, all the gas accreted onto a galaxy is
either ejected or forms stars.  This assumption is crude because galaxies do
not turn all the accreted gas instantaneously into stars.  However, even in
spiral galaxies, typical gas fractions (cold gas divided by cold gas plus
stars) rarely exceed $\sim 10-20\%$ (except for dwarf irregulars and blue
compact dwarfs, see e.g. \citealp{mcgaugh_etal10}).
Therefore, the error on the galaxy stellar mass that one makes by
assuming that the entire galaxy mass is in stars is consistent with the other
uncertainties of our model at all but the lowest masses.  
When we substitute 
$m_{\rm wind}=(v_{\rm SN}/v_{\rm circ})^2\widetilde m_{\rm stars}$ into 
$m_{\rm wind}+\widetilde m_{\rm stars}=m_{\rm accr}$, we find
\begin{equation}
\widetilde m_{\rm stars}={v_{\rm circ}^2\over v_{\rm circ}^2+v_{\rm
 SN}^2}\,m_{\rm accr}.
\label{mSN}
\end{equation}

\subsubsection{Synthesis}
\label{secsynth}

By putting together Eqs.~(\ref{mshock}), (\ref{mreion}), and (\ref{mSN}), we
can write a simple but physically motivated equation for the \emph{model}
stellar mass (in the absence of mergers) of
the central galaxy of a halo of mass $M_{\rm halo}$ in the absence of
merging:
\begin{equation}
\widetilde m_{\rm stars}=
{v_{\rm circ}^2-v_{\rm reion}^2\over v_{\rm circ}^2 + v_{\rm SN}^2}\,
{f_{\rm b}\,M_{\rm halo}\over 1+M_{\rm halo}/M_{\rm shock}},
\label{mstars}
\end{equation}
where 
\begin{eqnarray}
v_{\rm circ}&\equiv&v_{\rm circ}(M_{\rm halo},z) \nonumber \\
&=& \left [{\Delta(z)\over 2}\right]^{1/6} [G\,H(z)]^{1/3}\,M_{\rm halo}^{1/3}
\ .
\label{vcirc}
\end{eqnarray}
Had we taken into account that some gas is neither 
blown away by the supernovae nor converted into stars, then mass conservation  
 would have given $m_{\rm wind} + \widetilde m_{\rm stars}/(1-g) = m_{\rm accr}$, where $g$ is the
fraction of galaxy baryons in the form of remaining gas. Hence,
equation~(\ref{mstars}) would become
 \begin{equation}
\widetilde m_{\rm stars}=
{v_{\rm circ}^2-v_{\rm reion}^2\over v_{\rm circ}^2/(1-g) + v_{\rm SN}^2}\,
{f_{\rm b}\,M_{\rm halo}\over 1+M_{\rm halo}/M_{\rm shock}} \ .
\label{mstars2}
\end{equation}

At each timestep, the galaxy stellar mass is assigned according to
equations~(\ref{mstars}) and (\ref{vcirc}), unless this produces a lower
stellar mass than at the previous timestep.  

Equation~(\ref{mstars}) should
be seen more as an empirical fitting formula than as a physical model,
although the arguments that we have presented above provide some physical
justification for it. We are neglecting a host of other physical processes
such as the conditions for the formation of clumpy molecular clouds from cold
atomic gas, and mechanisms operating inside halos such as ram pressure
stripping, conduction, magnetic fields, cosmic rays, etc., because we wish to
build the simplest model that provides an adequate fit to the galaxy mass
function of galaxies and thus allows the study of the mass buildup of galaxies.

The key point of this article is not to present a new model for processes such as 
shock heating, cooling, star formation, and feedback (SAMs do that more satisfactorily),
but rather to estimate the role of mergers in the galaxy growth. For this we need a prescription
to associate a galaxy mass to a halo mass at each redshift and this is what Equation~(\ref{mstars}) does.
What is more important is that we compute galaxy mergers accurately and this is where our model is state-of-the-art
(Section~\ref{sec:galaxy_mergers}).

\begin{figure}
% GAM SM: sfevsMmodel sfevsM
\centering
\includegraphics[width=\hsize]{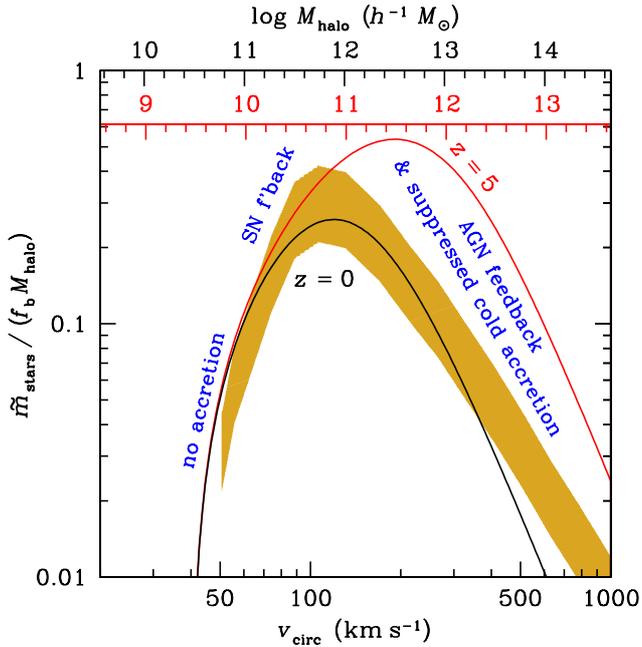}
\caption{Star formation efficiency versus halo circular velocity and 
mass at two different epochs in our toy model (eq.~[\ref{mstars}]). 
This figure has been plotted for the parameters of
Table~\ref{bestfit} that produce the best fit to the stellar mass function at
$z=0$ (see Sect.~\ref{sec:bestfit}). 
The \emph{upper (black)} mass scale is for 
$z$=0, while
the \emph{lower (red)} mass scale is for 
$z$=5.
The \emph{gold shaded region} shows the relation predicted by abundance
matching by \cite{GWLB10}.
\label{sfevsM}}
\end{figure}

Figure~\ref{sfevsM} 
illustrates how, in our toy model (eq.~[\ref{mstars}]), 
the efficiency with which halo mass
growth results in stellar mass growth depends on halo mass.
This efficiency has been plotted, at two different redshifts, 
for the parameter set that gives the best fit to the $z\!=\!0$
galaxy mass function (see Sect.~\ref{sec:bestfit} and Table~\ref{bestfit}
below). The star formation efficiency is maximum at $M_{\rm
  halo}\sim 10^{11}-10^{12}\,h^{-1}\,M_\odot$, decreasing sharply at lower
masses and more mildly at higher masses.  Figure~\ref{sfevsM} shows that in a
$z$=0 halo, $<20\%$ of the baryonic mass is expected to be in stars.  The
fraction of baryons that may be locked up in stars can reach 50\% at high
redshift, where feedback is less important because the circular velocity
thresholds of equation~(\ref{mstars}) correspond to lower mass thresholds
(eq.~[\ref{vcirc}], noting that $\Delta(z)$ and especially $H(z)$ increase
with $z$).
Our model produces a $z\!=\!0$ stellar mass vs. halo mass relation very similar to that
obtained by \cite{GWLB10} with the abundance matching technique, except that
our toy model (eq.~[\ref{mstars}]) underestimates the stellar mass at large
halo masses relative to the abundance matching solution. This will be
compensated by the effects of galaxy mergers (see eq.~[\ref{M1new}]) as we
shall see in Sect.~\ref{results} (Fig.~\ref{fig:mstarvsMhalo}) below.

\begin{figure}
%\centerline{\hbox{\psfig{figure=fig1.eps,width=0.9\hsize,angle=0}}}
\centering
\includegraphics[width=0.9\hsize]{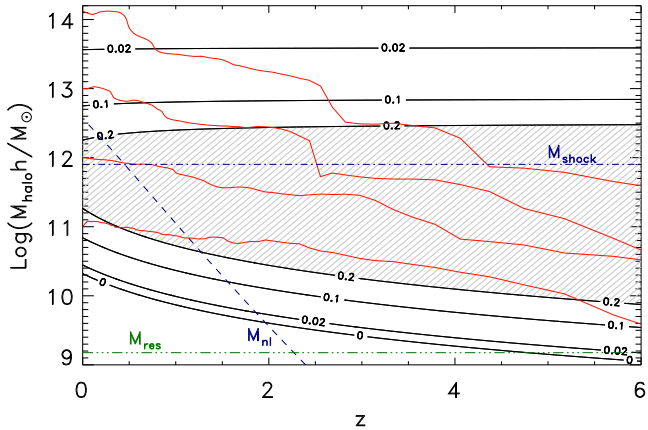}
\caption{Growth of halo mass and efficiency of star formation.  The
  \emph{contours} show the efficiency of star formation, $m_{\rm stars}^{\rm
    accr}/(f_{\rm b}\,M_{\rm halo})$, as a function of redshift and halo
  mass, for the set of parameters that fit best the galaxy mass function
  (Sect.~\ref{sec:bestfit} and Table~\ref{bestfit}).  The \emph{horizontal dash-dotted
    line} shows the critical halo mass for shock heating, $M_{\rm
    halo}=M_{\rm shock}$.  Baryons make stars efficiently only in the
  horizontal band of the diagram shown by the \emph{shaded region}.  The
  \emph{red irregular lines} show the mass growth of four haloes, whose
  present masses are about $10^{11}$, $10^{12}$, $10^{13}$ and
  $10^{14}h^{-1}M_\odot$.  They are backwards tracks for the most massive
  progenitors.  The \emph{diagonal dashed line} shows the characteristic mass
  scale $M_{\rm nl}$ on which density fluctuations become non-linear at
  redshift $z$, i.e. where $\sigma(M_{\rm nl},z)=1.68$.  The \emph{horizontal
    green line} shows the mass resolution $M_{\rm res}=1.5\times
  10^9h^{-1}M_\odot$ of the $N$-body simulation.  }
\label{fig:sfe}
\end{figure}

\Fig{sfe} shows how these processes quench star formation
during the growth of a dark matter halo.  The contours show $m_{\rm
  stars}/(f_{\rm b}\,M_{\rm halo})$ as a function of halo mass and redshift
(for the best-fit parameters used for Fig,~\ref{sfevsM}, 
see  Sect.~\ref{sec:bestfit} and Table~\ref{bestfit}). 
The term in $v_{\rm circ}$ in Eq.~(\ref{mstars}) suppresses galaxy
formation in low-mass haloes, while the cut-off at $M_{\rm shock}$ suppresses
galaxy formation in high-mass haloes.  The result is a \emph{galaxy formation
  zone} at $10^{10}M_\odot\lsim M_{\rm halo}h\lsim 10^{12.5}M_\odot$. Galaxy
formation begins and ends when a halo moves in and out of the galaxy
formation zone, respectively.  The four red curves in \Fig{sfe} show the mass
growth with redshift of four haloes whose final masses at $z$=0 are roughly
$10^{11}$, $10^{12}$, $10^{13}$ and $10^{14}h^{-1}M_\odot$. The haloes that
are more massive enter and exit the galaxy formation zone at an earlier
epoch. Since the most massive haloes contain the most massive galaxies, this
explains why the most massive galaxies contain the oldest stellar populations
(\emph{archaeological downsizing}, \citealp{cattaneo_etal08}).

The dashed curve in \Fig{sfe} shows the characteristic mass $M_{\rm nl}$ of
density fluctuations that become non-linear at redshift $z$, i.e.  the
characteristic mass scale on which the cosmic variance of linearly
extrapolated primordial density fluctuations is equal to $\delta_{\rm
  c}\simeq 1.68$ at redshift $z$.  Here $\delta_{\rm c}$ is the density
contrast of a top hat fluctuation that the linear theory predicts at the time
when the fluctuation actually collapses.  The value of $\delta_{\rm c}$
depends very weakly on $\Omega_{\rm m}$ and $\Omega_{\rm \Lambda}$.  The
redshifts at which $M_{\rm nl}$ comes in and out of the galaxy formation
zone mark the beginning and the end of the main epoch of galaxy formation in
the Universe.

Similar simple toy models were
independently constructed by D. Croton (private communication) and by
\cite{bouche_etal10}, although in their models the low-mass
cut-off does not depend on redshift, while in our model it is the circular
velocity that is fixed, so the minimum mass decreases with increasing
redshift (see \Fig{sfe}).

\subsection{Tidal stripping}
\label{tidal_str}

The dynamical friction time given by eqs.~(\ref{tdf}), (\ref{Coulomb}) and
(\ref{jiang}), which was calibrated by \cite{jiang_etal08} on hydrodynamical
cosmological simulations, implicitly incorporates the effects of tides, which
strip the secondaries to lower masses as they orbit through the primary, on
the dynamical friction time. However, knowledge of the stellar mass at a
given time is necessary to compare to the observed galaxy mass function, so
we need to estimate the time evolution of the stellar mass caused by tidal
stripping.

Tidal forces are known to strip galaxy haloes very efficiently once they
penetrate larger ones (e.g. \citealp{Ghigna+98}), and each pericentric
passage generates more mass loss \citep{Hayashi+03}. But tides also affect,
to a lesser extent, the more bound stellar material. For example,
\cite{Klimentowski+09} ran simulations of a low-mass high-resolution spiral
galaxy around the fixed potential of the Milky Way and found that over 5
orbits, while the dark matter mass was reduced by a factor 100, the stellar
mass was reduced by a factor 10, meaning that a fraction of $\eta_{\rm strip}
= 1-(1/10)^{1/5}=0.37$ of the stellar mass was lost at every pericentric
passage.  This tidally stripped stellar mass should form what is known as the
stellar halo in the case of a galaxy and the intracluster light in the case
of a cluster.

According to the simulations of \citeauthor{Klimentowski+09}, the mean mass
loss $\eta_{\rm strip}$ is roughly constant for every pericentric passage, so
the importance of tidal stripping depends on the number of orbital
revolutions that a galaxy makes from when its orbit starts decaying to when
either the galaxy merges with the central galaxy of its parent halo or the
simulation ends.  In the former case, the total time span during which a
galaxy is stripped is given by the time it takes to decay by dynamical
friction from the virial radius to the centre (Eq.~\ref{tdf} with $r$ and
$v_{\rm circ}$ taken at the virial radius).  
Taking $t_{\rm orb}\sim 2\pi
r/v_{\rm circ}$ for the orbital time, we find
\begin{equation}
\tau = {t_{\rm df}\over t_{\rm orb}}= {A(\epsilon)\over 2\pi}\, {r_{\rm vir}v_{\rm
    circ}^2\over{\rm G}M_{\rm sat}\,\ln\Lambda} ={A(\epsilon)\over 2\pi}
\,{M_{\rm halo}^{\rm host}/ M_{\rm sat} \over \ln (1 + M_{\rm halo}^{\rm
    host}/ M_{\rm sat})}\ .
\label{strip}
\end{equation}
The normalisation, $A(\epsilon)$, of the dynamical friction time is computed
using the formula in \citet{jiang_etal08}, which fits the results of SPH
simulations and accounts for the tidal stripping of sub-structures.
Eq.~(\ref{strip}) implies that, for a fixed value of $\eta_{\rm strip}$, the
importance of tidal stripping increases with $M_{\rm halo}^{\rm host}/M_{\rm
  sat}$.

\subsection{Summary}

In summary, the final equation for the galaxy stellar at any given time is
\begin{equation}
\label{M1}
m_{\rm stars} = {\rm max}\left\{  \begin{array}{l}
\displaystyle
\widetilde m_{\rm stars}[M_{\rm
    halo},v_{\rm circ}(M_{\rm halo},z)]\ ,\\
\\
\displaystyle
\sum_{\rm prog} m_{\rm stars}^{\rm (prog)}(1-\eta_{\rm strip})^\tau
\ ,
\end{array}
\right.
\label{updatemass}
\end{equation}
where $\widetilde m_{\rm stars}$ and $\tau$ are given in
equations~(\ref{mstars}) and (\ref{strip}), respectively, with
$\tau=0$ for the main progenitor.
Equation~(\ref{updatemass}) contains four free parameters ($v_{\rm reion}$,
$v_{\rm SN}$, $M_{\rm shock}$, and $\eta_{\rm strip}$), 
the first three of which enter the model through the function $\widetilde
m_{\rm stars}$ of equations~(\ref{mstars}) and (\ref{vcirc}).
The fourth parameter $\eta_{\rm strip}$ affects
the final galaxy masses by controlling the efficiency of tidal stripping,
but, as we shall see, 
$\eta_{\rm strip}$ has comparatively little bearing on
our results.
There are no free parameters in the calculation of the halo and thus galaxy
merger rate.

\section{Results}
\label{results}

\subsection{Best-fit model and effects of physical processes}
\label{sec:bestfit}

We start with the  set of our 4 parameters, $v_{\rm reion}$,
  $v_{\rm SN}$, $M_{\rm shock}$, and $\eta_{\rm strip}$, that provide a
  best-fit to the galaxy stellar mass function deduced by \citet{bell_etal03}
  from the SDSS observations,
and then consider how the predictions vary when the assumptions are changed.

The stellar masses of galaxies derived by \citet{bell_etal03} to compute the
SDSS mass function
depend on the Hubble constant used to compute galaxy
luminosities and on the stellar initial mass function (IMF).  The masses
inferred from observed luminosities scale as $h^{-2}$ while simulated masses
scale as $h^{-1}$.  As it is not possible to make the comparison between
model and observations completely independent of $h$, we convert the
observationally determined masses in units of $h^{-1}M_\odot$ by assuming
$h=0.7$.  \citet{bell_etal03} compute the stellar mass function using a diet
Salpeter IMF \citep{bell_dejong01}.  To convert to a \citet{salpeter55} or a
\citet{kroupa02} IMF, one should add a correction of $+0.15$ and $-0.1\,$dex,
respectively (see \citealp{bell_etal03}).

\begin{table}
\begin{center}
\caption{best-fit parameters corresponding to the full model in \Fig{MFs}.}
\label{bestfit}
\begin{tabular}[h]{ l | l }
\hline			
 $v_{\rm reion}$ & \ \,40${\rm\,km\,s}^{-1}$ \\
 $v_{\rm SN}$    & 120${\rm\,km\,s}^{-1}$ \\
 $M_{\rm shock}$ & $8\times 10^{11}h^{-1}M_\odot$ \\
 $\eta_{\rm strip}$ & 0.4 \\
\hline  
\end{tabular}
\end{center} 
\end{table}

\begin{figure}
\noindent
%\begin{minipage}{8.4cm}
 \centerline{\hbox{
     \psfig{figure=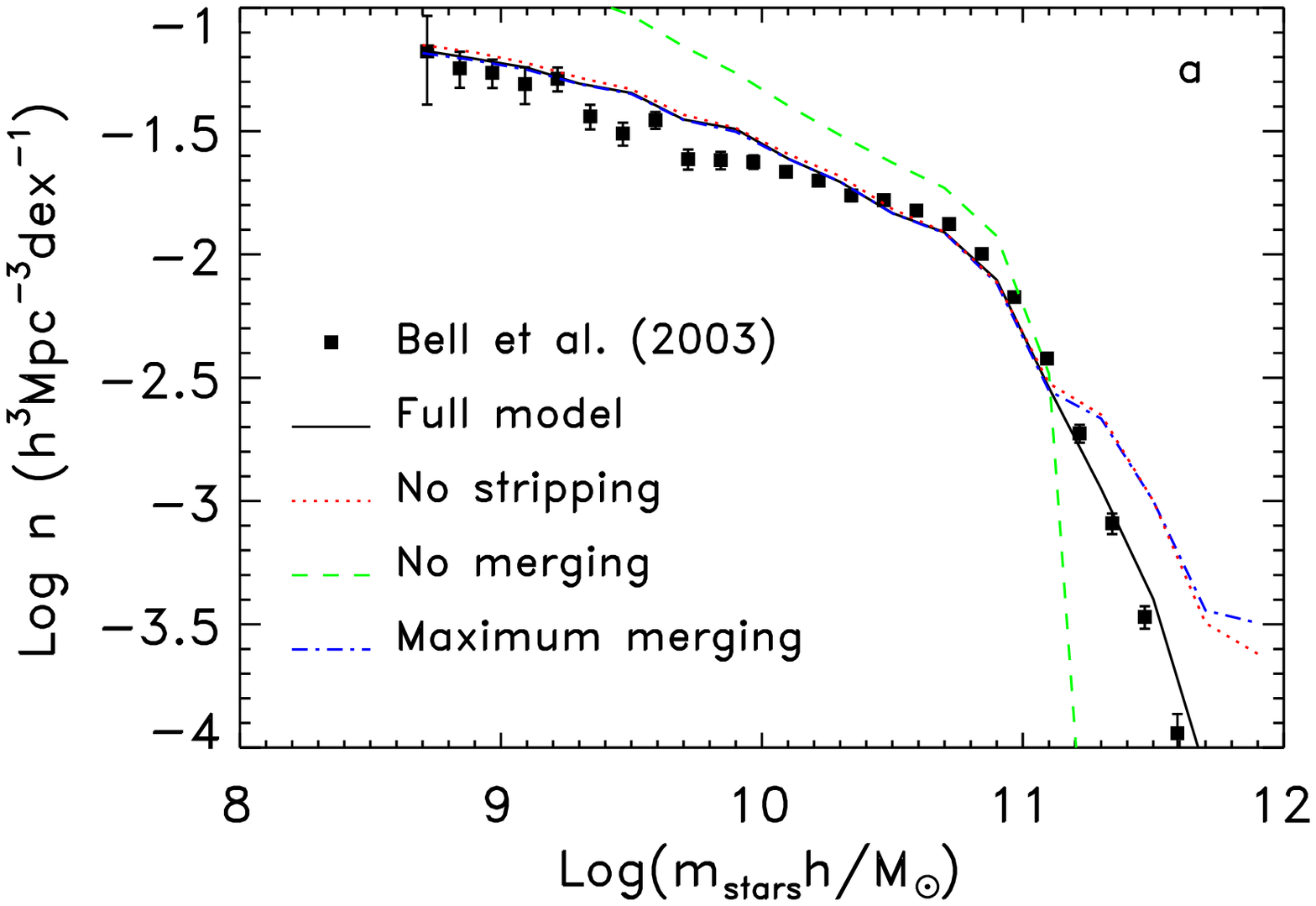,width=\hsize,angle=0}
 }}
%\end{minipage}\    \ 
%\begin{minipage}{8.4cm}
 \centerline{\hbox{
     \psfig{figure=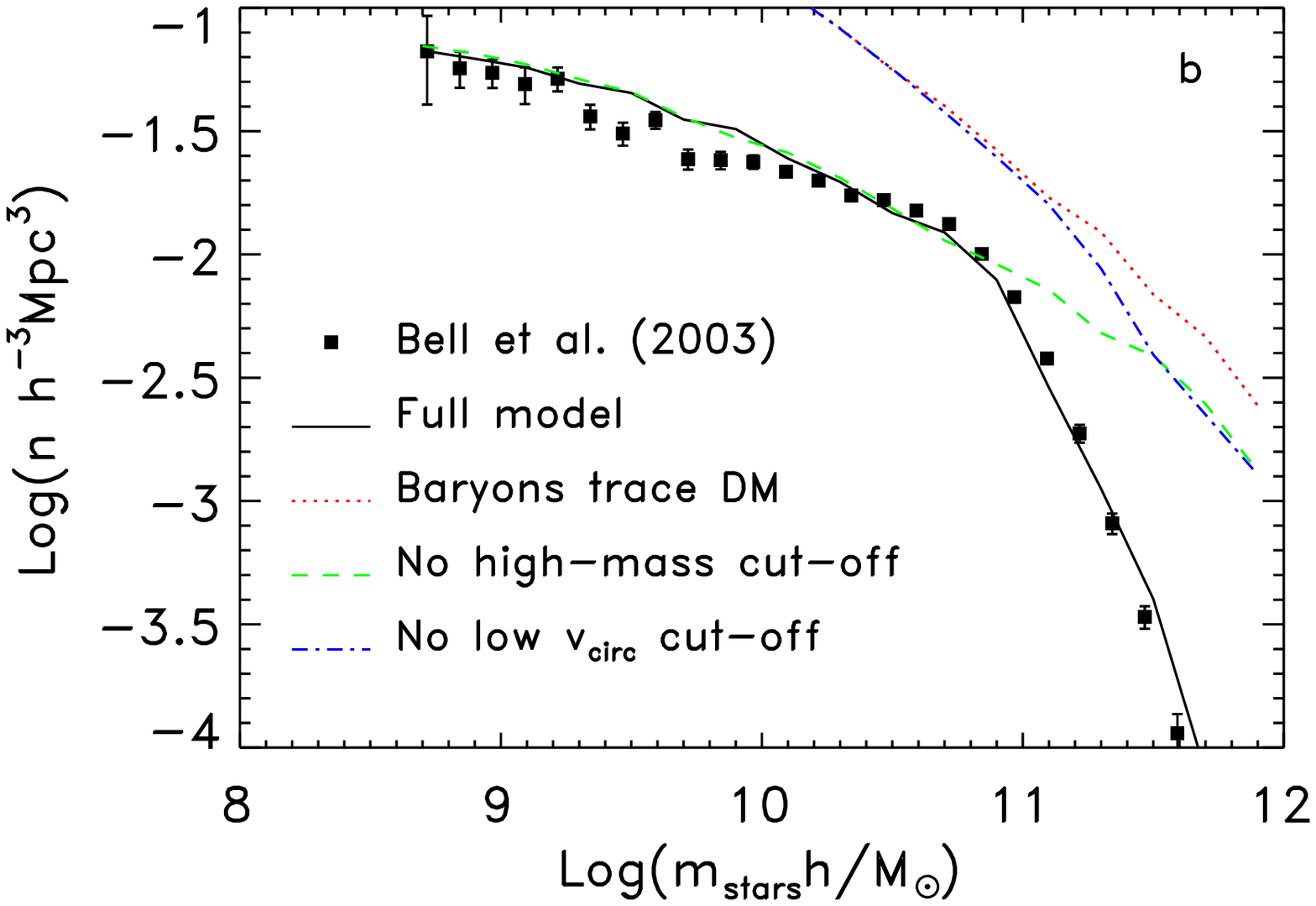,width=\hsize,angle=0}
 }}
%\end{minipage}\    \ 
\caption{Stellar mass function of galaxies at $z$=0.  The \emph{filled
    squares} show the galaxy mass function determined observationally by Bell
  et al. (2003).  The \emph{solid line} shows our best-fit model (the `full'
  model, with parameter values listed in Table~\ref{bestfit}).  The fit
  assumes $h=0.7$ because observationally inferred masses scale as $h^{-2}$.
  (a) Variations of the stellar mass function for different assumptions on
  merging and stripping: best-fit full model (\emph{black solid line}), no
  tidal stripping ($\eta_{\rm strip}=0$, \emph{red dotted line}), no mergers
  at all (\emph{green dashed line}), and maximum merging 
(\emph{blue dash-dotted line}).  (b) Variations of the stellar mass function for
  different assumptions on the baryonic physics: best-fit full model
  (\emph{black solid line}, same as in panel a), baryons trace dark matter ($m_{\rm stars} =
  f_{\rm b}\,M_{\rm halo}$, \emph{red dotted line}), no high-mass cutoff
  ($M_{\rm shock}\to\infty$, \emph{green dashed line}), and no low circular
  velocity cutoff ($v_{\rm SN} = v_{\rm reion} = 0$, \emph{blue dash-dotted line}). 
In both panels, all variations start from the best-fit
  parameters of the full model except for the fact that we use $\eta_{\rm
    strip}=0$ in all models apart from the full one.  
}
\label{fig:MFs}
\end{figure}

The solid lines in the two panels of \Fig{MFs} indicate that
our best-fit model (Table~\ref{bestfit})
matches well the
\citet{bell_etal03} mass function at $10^{8.7}M_\odot<h\,m_{\rm
  stars}<10^{11.6}M_\odot$.  Different determinations of the stellar mass
function of galaxies (e.g. \citealp{baldry_etal08, yang_etal09}) would lead
to slightly different
best-fit parameters.  Reasonable modifications to the IMF
  will also yield slightly different best-fit parameters.  However, the conclusions of our
article  are robust to these uncertainties, as they do not depend on the precise
values of the model parameters.
We also note that $v_{\rm SN}=120\,\rm
km\,s^{-1}$ is close to the value of $100 \, \rm km \, s^{-1}$ derived by
\cite{dekel_silk86}. It implies a supernova efficiency of $\sim 1\%$ if there
is one $10^{52}\,$erg supernova every $100\,M_\odot$ of star formation and
the escape speed is $\sqrt 3\,v_{\rm circ}$, that is, the required wind
kinetic energy is $\sim 1\%$ of the supernova energy.

With $v_{\rm reion}=40 \, \rm km \, s^{-1}$ our model cannot form galaxies
below this halo circular velocity. One may argue that galaxies are known with
lower maximum rotation velocities. Indeed, since our model starts at roughly the
epoch of reionisation, we will be missing the galaxies of lower mass that
form before reionisation
(see, e.g., \citealp{MTCT10}). It turns out that the precise value of $v_{\rm
  reion}$ has little effect on the conclusions of our paper, because of our
finite galaxy mass resolution. Moreover, relative to their halo masses, their
stellar masses are 
extremely low, hence they contribute little to the growth of more massive
galaxies.

The incorporation of $\eta_{\rm strip}$ reduces the stellar masses of
satellite galaxies at each timestep.  Comparison with the case $\eta_{\rm
  strip}=0$ (\Fig{MFs}a, dotted line) shows that this effect is negligible at
$m_{\rm stars}\lsim 10^{11}h^{-1}M_\odot$. 
Tidal stripping is, however, important at higher masses, where the galaxy
mass function is predicted to be shifted toward high masses by $\sim
0.2\,$dex in mass compared to the \citet{bell_etal03} mass function if
satellites are not stripped of part of their mass before they merge into the
galaxies that populate this part of the mass function.  This discrepancy may
also be due to systematic measurement errors in the data, since many
massive ellipticals extend outside the photometric aperture used to measure
their luminosities.  \citet{lauer_etal07} find errors of up to a magnitude in
the SDSS luminosities of such galaxies and these are the same data that
anchor the high end of the \citeauthor{bell_etal03} mass function in
\Fig{MFs}.  We shall elaborate on this in the Discussion
(Section~\ref{sec:discuss-tidal}). 

Assuming that every halo merger immediately results in a galaxy merger causes
the most massive galaxies to grow even larger (dash-dotted line in \Fig{MFs}a) but
the difference between the dash-dotted line and the dotted line is fairly
minor suggesting that most halo mergers do result in galaxy mergers.  The
difference between the two curves in the highest mass bin disappears if the dash-dotted
line is computed not for all halo mergers, but for those of bound
halo-satellite systems.

The dashed line in \Fig{MFs}a corresponds to the extreme opposite assumption
that whenever a new halo appears a new galaxy is created but there are no
galaxy mergers at all. This assumption predicts too many low-mass galaxies
and too few high-mass galaxies with respect to the observations.

Both the dotted line and the dash-dotted line in \Fig{MFs}a assume models
without tidal stripping (so does the dashed line).  In both cases the cosmic
stellar mass density obtained by integrating the galaxy stellar mass function over
all masses exceeds the observational value inferred from the
\citet{bell_etal03} mass function. This suggests that stripping is necessary,
since, in our model,
changing the merging rate does not affect the cosmic stellar mass
density.

Having seen how the mass function depends on the capture and the probable
stripping of satellites, we now examine how it depends on our assumptions
on the quiescent growth of galaxies in isolated haloes (\Fig{MFs}b).  We
compare the full model with three unrealistic simpler models, which are,
however, useful for illustrative purposes.  They are: i) a model in
which the stellar mass grows proportionally to the mass of the dark matter
($m_{\rm stars}=f_{\rm b}\,M_{\rm halo}$; dotted line); ii) a model
with the term in $v_{\rm circ}$ but without the cutoff at high masses (dashed
line); and iii) a model with the cutoff at high masses but not the term in
$v_{\rm circ}$ (dash-dotted line).

\Fig{MFs}b illustrates the well known fact that supernova and reionisation
feedback processes are essential to reconcile the observed galaxy mass
function with the halo mass function of a cold dark matter Universe (compare
the solid line and the dash-dotted line). The suppression of star formation
at $M_{\rm halo}>M_{\rm shock}$ is necessary to reproduce the break in the
galaxy stellar mass function at $\sim 10^{11}M_\odot$ (compare the solid line
and the dashed line).  It is the combination of these low and high mass
cut-offs that allows the best-fit model to reproduce the normalisation and
the characteristic break of the galaxy stellar mass function.  
\Fig{MFs}b shows that the low $v_{\rm circ}$ cut-off  
is important, not only for the low-mass end, but also
for the high-mass end of the galaxy mass function, because the high-mass
cut-off alone is not enough if the building  
blocks of giant ellipticals have been allowed to grow in an uncontrolled
fashion before crossing the  
critical mass.\footnote{This result agrees with what N. Katz finds from hydrodynamic simulations
(private communication).} 
We also find that
the quenching of high mass galaxies by shock heating is much more effective
than the reduction of their masses by tidal stripping, as can be seen by
comparing Figs.~\ref{fig:MFs}a (red dotted curve) and \ref{fig:MFs}b (green
dashed curve).

\begin{figure}
%\centerline{\hbox{\psfig{figure=Madau.eps,height=6.6cm,angle=0}}}
\centerline{\psfig{figure=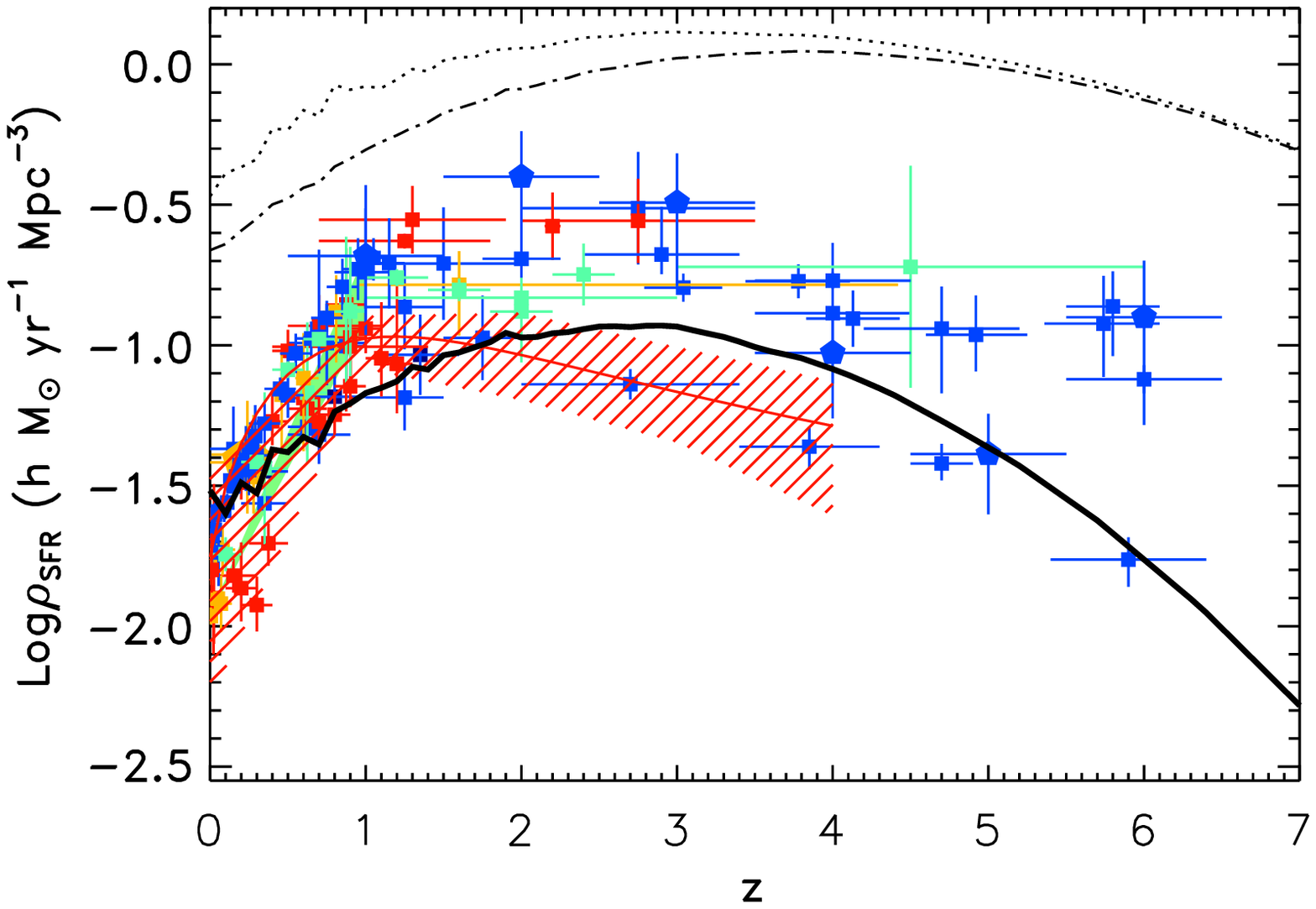,width=\hsize,angle=0}}
\caption{The evolution of the cosmic SFR density with redshift in the
  observed Universe (\emph{points with error bars} and \emph{hatched region}) and in our model
  (\emph{thick black solid line}; the \emph{dotted} and \emph{dotted-dashed lines}
correspond to the two models where baryons trace dark matter and without a low circular
velocity cutoff, respectively, see Fig.~\ref{fig:MFs}b).  The \emph{blue squares} are UV data
  \citet{giavalisco_etal04,wilson_etal02,massarotti_etal01,sullivan_etal00,steidel_etal99,cowie_etal99,treyer_etal98,connolly_etal97,baldry_etal05,schiminovich_etal05,wolf_etal03,arnouts_etal05,bouwens_etal03a,bouwens_etal05,bunker_etal04,ouchi_etal04}.
  The \emph{blue pentagons} are Hubble Ultra Deep Field estimates
  \citep{thompson_etal06}.  The \emph{red squares} are H$\alpha$, H$\beta$
  and OII data
  \citep{hanish_etal05,perezgonzalez_etal03,tresse_etal02,hopkins_etal00,moorwood_etal00,sullivan_etal00,glazebrook_etal99,yan_etal99,tresse_maddox98,gallego_etal95,pettini_etal98,teplitz_etal03,gallego_etal02,hogg_etal98,hammer_etal97}.
  The \emph{orange squares} are 1.4$\,$GHz data
  \citep{mauch_sadler07,condon_etal02,sadler_etal02,serjeant_etal02,machalski_godlowski00,haarsma_etal00,condon89}.
  The \emph{orange pentagon} at $z\simeq 0.24$ is X-ray data
  \citep{georgakakis_etal03}.  The \emph{green squares} are infrared
  \citep{perezgonzalez_etal05,flores_etal99} and sub-mm \citep{barger_etal00}
  data.  The \emph{green hatched region} is the far infrared ($24\,\mu m$)
  SFR history from \citet{lefloch_etal05}.  
  Most of
  them were taken from the compilations in \citet{hopkins04} and
  \citet{hopkins_beacom06}.  The \emph{red curve}, $\dot \rho_{\rm stars} =
  (0.014+0.11\,z)\,h/[1+(z/1.4)^{2.2}](h/0.7)\,M_\odot{\rm\,Mpc}^{-3}$, is
  the cosmic SFR derived by \citet{wilkins_etal08} by taking the derivative
  of the cosmic stellar mass density, i.e. from the variation of the observed
  stellar mass functions with $z$ rather than from star formation rate
  measurements. The \emph{red shaded area} shows the margins of uncertainty
  around the red curve.  All observational data have been corrected
  ($-0.15\,$dex) to the diet Salpeter IMF assumed by Bell et al. (2003), on
  whose mass function we have calibrated our model.}
\label{fig:cosmicSFR}
\end{figure}

\begin{figure*}
% GAM SM: mvsMwfmwet2 mvsM 1
\centering
\includegraphics[angle=-90,width=\hsize]{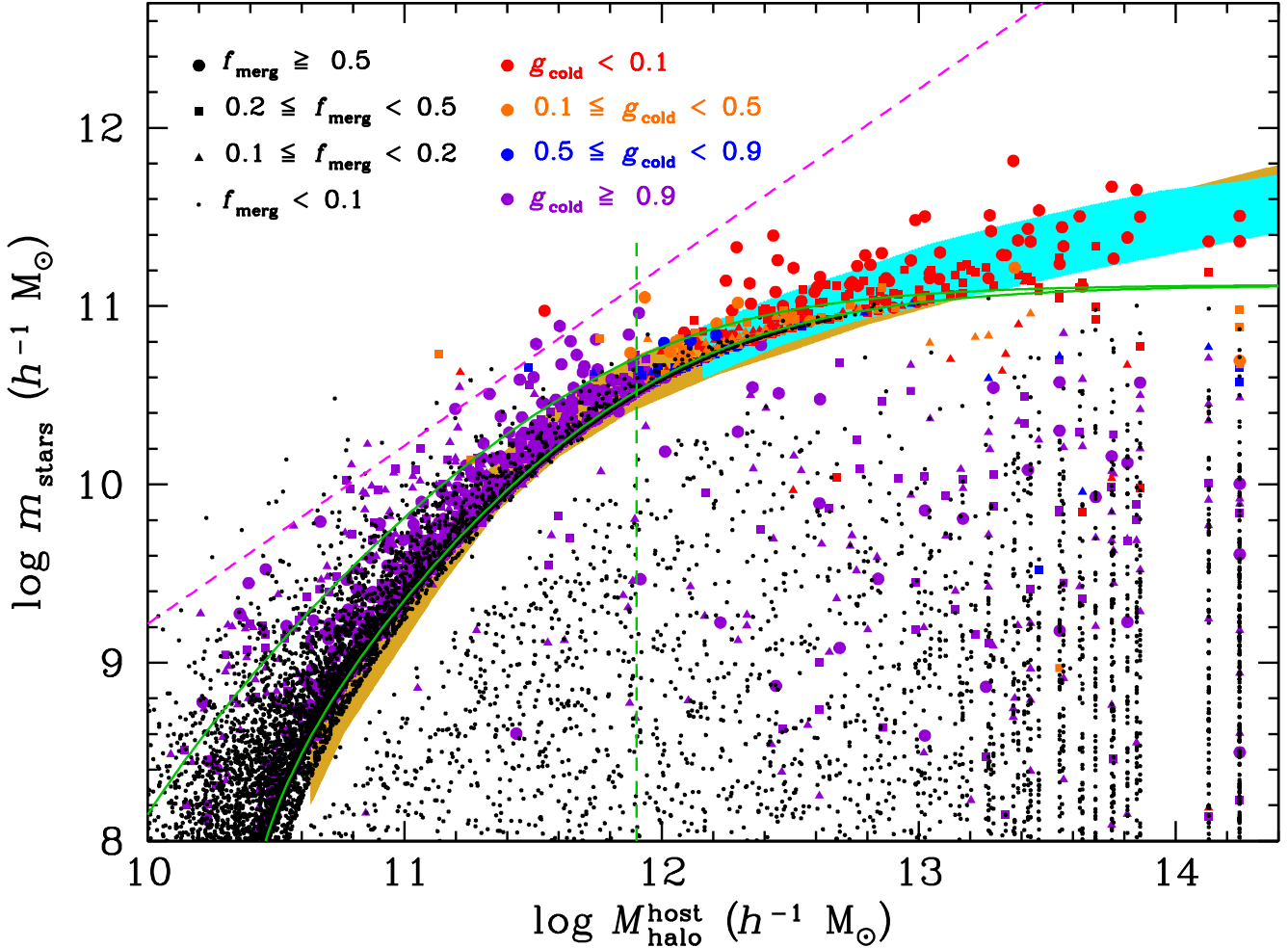}
\caption{Galaxy stellar mass vs. host halo mass at $z$=0 for the set of
  parameters (Table~\ref{bestfit})  that fit best  the galaxy stellar mass function.  Each point is
  one of our simulated galaxies.  The symbol sizes and colours vary with the
  fraction $f_{\rm merg}$ of mass acquired in mergers, and the fraction
  $g_{\rm cold}$ of this mass acquired by cold-mode mergers (where the sum of
  the masses
  of the haloes of the merging galaxies is below $M_{\rm shock}$),
  respectively, according to the legend.  The \emph{vertical green line}
  corresponds to $M_{\rm halo}=M_{\rm shock}$, while the \emph{oblique
    magenta line} indicates $m_{\rm stars} = f_{\rm b}\,M_{\rm halo}$.  The
  \emph{green curves} show the stellar masses of galaxies acquired through the
  accretion channel (eq.~[\ref{mstars}] with best-fit parameters from
  Table~\ref{bestfit}) at $z$=0 (\emph{lower}) and $z=3$ (\emph{upper
    curve}).  The \emph{shaded cyan} and \emph{gold regions} correspond to the observations
  from the SDSS (Yang et al. 2009), and the halo-galaxy abundance matching of
  \citet{GWLB10}. For the latter, the halo masses were shifted by $\log_{10}
  h$ to conform to our units. Also, the stellar masses were shifted by
  $-\log_{10} h$ and $\log_{10} h$ for \citeauthor{yang_etal09} and
  \citeauthor{GWLB10}, respectively, to take into account the different units
  with $h$. We also shifted the stellar masses by an additional factor of
  +0.1 and $+0.15\,$dex to pass from the Kroupa and Chabrier IMFs,
  respectively assumed by \citeauthor{yang_etal09} and \citeauthor{GWLB10} to
  the diet Salpeter IMF assumed by Bell et al. (2003), on whose mass function
  we have calibrated our model. 
}
\label{fig:mstarvsMhalo}
\end{figure*} 

Figure~\ref{fig:cosmicSFR} shows the evolution of the cosmic SFR density.
There are two ways to compute the cosmic SFR density: by observations (or SAM
simulations) of the total rate of star formation, or by taking the time
derivative of the total mass integrated over the galaxy stellar mass
function.
Since our model does not incorporate the concept of SFR, we adopt the latter
method, which we show as the black lines (solid for our best-fit model).
The agreement between our
best-fit model and the data (especially those derived with the second method, e.g.  \citealp{wilkins_etal08}, red curve) is reasonable
given how simple our model is\footnote{The discrepancy in
  Figure~\ref{fig:cosmicSFR} between the directly measured SFR densities
  (symbols with error bars) and the evolution of the SFR density computed
from the time derivative of the integrated galaxy stellar mass function (red curve and shaded
  area) is an open observational problem.}.  

\subsection{Central and satellite galaxies: the role of the environment}
\label{Central and satellite galaxies}

\Fig{mstarvsMhalo}
provides a deeper insight of what happens inside our model. We have plotted all our
simulated galaxies, at $z=0$, in an $m_{\rm stars}$ vs. $M_{\rm halo}^{\rm host}$
diagram, where $M_{\rm halo}^{\rm host}$ is the mass of the largest halo in
which a galaxy is contained (so that if a small galaxy is located near a
larger one, itself within a group of galaxies, the host halo is that of the
group and not that of the larger galaxy).  
In other words, \Fig{mstarvsMhalo} constitutes a representation of the
Tully-Fisher (\citeyear{tully_fisher77}) scaling relation between stellar
mass in galaxies and their halo 
properties (where the halo is quantified here by the mass at the virial
radius instead of the maximum circular velocity).
The symbol with which a galaxy is
plotted has 
been sized and colour-coded according to the values of the fraction $f_{\rm merg}$ of
present-day mass acquired via mergers (symbol size)
and the fraction $g_{\rm cold}$
of the mass acquired by mergers that was acquired in 
`cold-mode' mergers (symbol colour), where `cold-mode' and `hot-mode' mergers are defined as follows.
A merger that takes place between the two N-body snapshots $s$ and $s+1$ is classified as being `cold-mode' if
the mass of the host halo of the merger remnant at $s+1$ is $M_{\rm halo}^{\rm host}<M_{\rm shock}$.
\Fig{mstarvsMhalo} only shows the best-fit model, corresponding to the solid line in \Fig{MFs}.

We clearly see two galaxy populations separated by an empty zone: a
population of central galaxies (galaxies for which $M_{\rm halo}^{\rm
  host}=M_{\rm halo}$), which follows a narrow curved band in the $m_{\rm
  stars}$ -- $M_{\rm halo}^{\rm host}$ plane, roughly following the stellar
masses predicted from the accretion channel (eq.~[\ref{mstars}])
and a population of satellite galaxies
($M_{\rm halo}<M_{\rm halo}^{\rm host}$), which are scattered over a broad
area of the diagram lying below the central galaxy relation.

It is important to note that the masses in the $y$-axis of \Fig{mstarvsMhalo} are
stellar masses. Neglecting the mass in cold gas
is reasonable for massive galaxies, but not at
low stellar masses ($\log h m_{\rm stars} < 8.7$, e.g. \citealp{zhang_etal09}), 
where the HI mass dominates the stellar mass. The inclusion of gas in our toy
model (eq.~[\ref{mstars2}]) with gas-to-star ratios given in
\citeauthor{baldry_etal08} (\citeyear{baldry_etal08}, and references therein)
leads to baryonic masses of low-mass 
galaxies that are nearly twice as large as inferred from
the stellar masses in \Fig{mstarvsMhalo}.

Central galaxies with $f_{\rm merg}\lsim 0.1$ tend to accumulate in a narrow
zone of the $m_{\rm stars}$ -- $M_{\rm halo}^{\rm host}$ diagram, which
appears as a black curve of points, and corresponds to the present-day
$m_{\rm stars}^{\rm accr}$ -- $M_{\rm halo}$ relation (lower green curve)
given by Eq.~(\ref{mstars}).  Its broadness is due to the higher value, for
given $M_{\rm halo}^{\rm host}$, of $v_{\rm circ}$ and therefore $m_{\rm
  stars}$ at higher redshifts, as can be seen from the two green curves in
Figure~\ref{fig:mstarvsMhalo}.  Galaxies that lie well above the critical
$z$=0 curve either live in haloes that have lost mass due to tidal stripping
(dots and triangles) or are galaxies that have grown above the relation by
mergers (filled circles).
Galaxies lying below the critical $z=0$ curve are the centrals of hosts
containing a non-negligible fraction of their mass in satellites, so that the
host halo mass is larger than the mass of the halo directly associated with
the galaxy.

Our general
$m_{\rm stars}$ -- $M_{\rm halo}^{\rm host}$ relation predicted for the
central galaxies of groups and clusters in \Fig{mstarvsMhalo} is highly
consistent with that found in the SDSS shown in Fig.~4 of \citet{yang_etal09}
(shaded cyan region in Fig.~\ref{fig:mstarvsMhalo}).  At lower masses, it is also
highly consistent with the relation that \cite{GWLB10} deduced (shaded gold
region in Fig.~\ref{fig:mstarvsMhalo}) by matching
the halo and subhalo mass function measured in the Millennium-II cosmological simulation
to the galaxy mass function deduced by \cite{li_white09}. This gives us
some confidence that our model of equation~(\ref{mstars}) 
provides a reasonable approximation of the true stellar -- halo mass relation.

\begin{figure}
% GAM SM: mfunsvsMandN10 mfunvsMandN10 mfunvsMandN10
\centering
\includegraphics[width=\hsize]{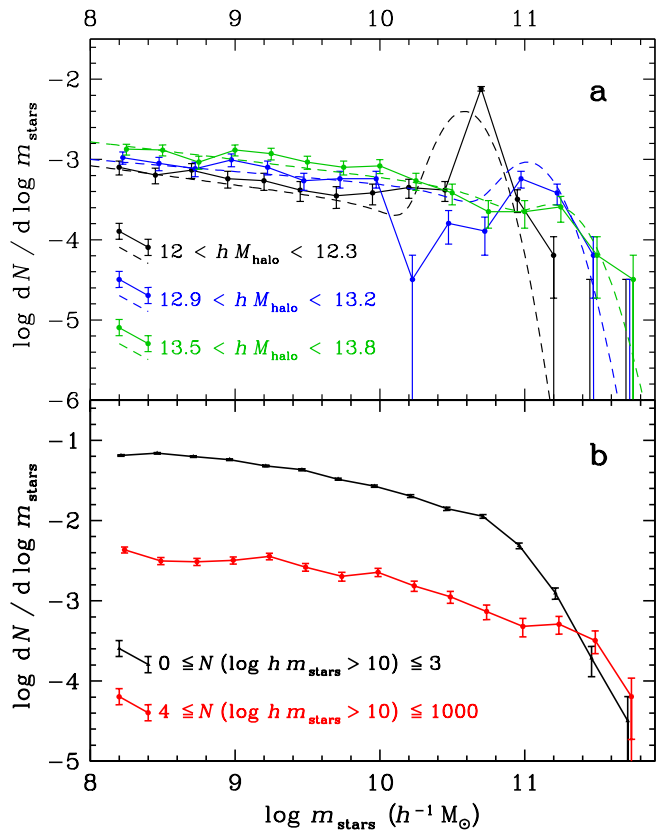}
\caption{The galaxy stellar mass function at $z$=0 split by environment, 
defined by host halo mass (\emph{top}) or group richness (number of
companions with $m_{\rm stars}>10^{10}M_\odot/h$ 
in the same host halo, \emph{bottom}).
The dashed curves in the top panel show the conditional mass function fits to
the SDSS data by Yang et al. (2009). 
The same $h$ and IMF corrections as in Fig.~\ref{fig:mstarvsMhalo} apply.
}
\label{masfuns}
\end{figure}

Figure~\ref{masfuns}a shows how different environments contribute to the
galaxy stellar mass function at $z$=0.  The galaxy mass function in haloes
with $10^{12}M_\odot<h M_{\rm halo}<10^{12.3}M_\odot$ contains a strong peak
at $m_{\rm stars}\sim 10^{10.7}h^{-1}M_\odot$, the characteristic mass of the
central galaxy of a halo in that mass range 
(Fig.~\ref{masfuns}a, black curve).  The contribution of satellites to the
galaxy mass function is at least an order of magnitude smaller than that of
the central galaxies.  In a halo with $10^{12.9}M_\odot<h M_{\rm
  halo}<10^{13.2}M_\odot$, the relative contribution of satellite galaxies is
higher but we still see a clear valley between central and satellite galaxies at
$m_{\rm stars}\sim 10^{10.4}h^{-1}M_\odot$, while the central galaxy peak has
moved up to $m_{\rm stars}\sim 10^{11.1}h^{-1}M_\odot$ (blue curve).  The
peak of the green curve is at a lower number density than the peak of the
blue curve, because there are less central galaxies of haloes with
$10^{12.9}M_\odot<h M_{\rm halo}<10^{13.2}M_\odot$ than there are central
galaxies of haloes with $10^{12}M_\odot<h M_{\rm halo} <10^{12.3}M_\odot$.
At $10^{13.5}M_\odot<h M_{\rm halo}<10^{13.8}M_\odot$, the satellite
population has become so numerous and the central galaxy peak has moved down
so much that the valley is no longer visible (green curve).  Thus, the more
massive the halo, the greater is the relative importance of the satellite population.

Similar bimodal (wide satellite + peaked central) galaxy stellar 
mass functions at given
halo mass are also seen in the analysis of SDSS galaxies
by \cite{yang_etal08} (their Fig.~2) and
\cite{yang_etal09} (their Fig.~4), with peaks and troughs 
occurring at very similar values of
$m_{\rm stars}$ (dashed and solid curves in Fig.~\ref{masfuns}a, respectively
for the SDSS and our model, where 
we used the same narrow bins  of halo mass as they did for
better comparison). The lack of galaxies with stellar masses just below those of central galaxies
(\Fig{mstarvsMhalo}) thus 
appears to 
be consistent with the two population (centrals + satellites) model (e.g., 
\citeauthor{yang_etal09}).

For observers, it is easier to define the environment of a galaxy in terms of
the number of companions above a luminosity limit than by its halo mass.  For
this reason, we show in Figure~\ref{masfuns}b how the galaxy stellar mass
function differs between rich and poor systems, where rich and poor are
defined by having respectively less than four or at least four companions
with $m_{\rm stars}>10^{10}h^{-1}M_\odot$.  Since we use two wide bins of
richness in Figure~\ref{masfuns}b, we cannot see peaks as we did in
Figure~\ref{masfuns}a.  However, there is an important difference between the
mass functions of poor and rich systems: poor systems display a strong lack
in massive galaxies relative to richer systems. In contrast, the low-end
slopes of the mass function is the same for poor and rich haloes.  In other
words, if one used the \citet{Schechter76} parametrisation, $f(m) \propto
m^\alpha \exp(-m/m_*)$, the exponential cutoff of the mass function, $m_*$,
would be much lower for the poor systems, while the faint-end slope, $\alpha$
would be independent of richness.

\subsection{The importance of mergers}
\label{sec:mergers}
\begin{figure}
%\begin{minipage}{8.4cm}
 \centerline{\hbox{
     \psfig{figure=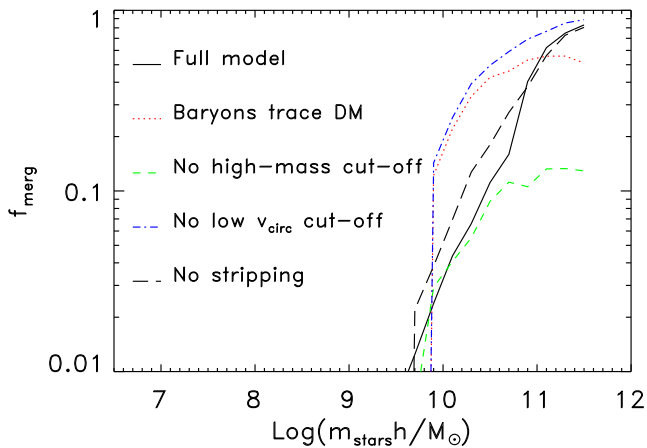,width=\hsize,angle=0}
 }}
%\end{minipage}\    \
\caption{Median fraction of the $z$=0 galaxy mass acquired via mergers rather
  than via gas accretion for various models:
 best-fit full model (\emph{black solid line}), 
baryons trace dark matter ($m_{\rm stars} = f_{\rm b}\,M_{\rm
  halo}$, 
\emph{red dotted line}),
no high-mass cutoff ($M_{\rm shock}\to\infty$, \emph{green short-dashed line}),
no low circular velocity cutoff ($v_{\rm SN} = v_{\rm reion} = 0$, 
\emph{blue dash-dotted line}),
and no tidal stripping ($\eta_{\rm strip}=0$, \emph{black long-dashed line}).
}
\label{fig:fmergers}
\end{figure}

\Fig{fmergers} shows the median value of the mass fraction, $f_{\rm merg}$,
acquired by mergers (instead of by gas accretion) in different mass bins for
the five models plotted in \Fig{MFs}a.  This median hides a large scatter, as
one can see from the distribution of $f_{\rm merg}$ at constant $m_{\rm
  stars}$ in \Fig{mstarvsMhalo}, but several statistically significant trends
emerge, as we shall now see.

Our standard model (black solid curve) shows that, for final galaxy masses
above $10^{11}\,h^{-1} M_\odot$, typical galaxies have acquired the bulk of
their mass by galaxy mergers, whereas at lower masses, typical
galaxies grow mainly by gas accretion.

Could the lack of mergers at low mass be a consequence of our limited mass
resolution?
One can compare our standard model to one where galaxies follow the haloes:
$m_{\rm stars}=f_{\rm b}\,M_{\rm halo}$ (red dotted curve).
The median fraction of mass acquired by mergers in the model where
baryons trace the dark matter 
decreases with decreasing final galaxy mass for 
$m_{\rm stars} \ga 10^{10.8} h^{-1} M_\odot$, which suggests possible effects
of mass resolution at lower final galaxy masses.
However, for $m_{\rm stars} > 10^{10.6} h^{-1} M_\odot$,
this decrease is much sharper in our standard model. 
In other words,
as one goes from the highest galaxy masses down to $m_{\rm stars} > 10^{10.6}
h^{-1} M_\odot$,
the median mass fraction acquired
by mergers decreases much faster in our standard model 
than in our model where baryons trace the
dark matter.
We thus conclude that the decreasing importance of galaxy mergers from 
$m_{\rm stars} = 10^{11.0} h^{-1} M_\odot$
to
$m_{\rm stars} = 10^{10.6} h^{-1} M_\odot$
is robust to the effects of our limited mass resolution.

Note that the host haloes of galaxies with  $m_{\rm stars}\sim
10^{10.5}h^{-1}M_\odot$ are well resolved.  
It is their merging histories that are not.
Neglecting all mergers with sub-resolution haloes ($M_{\rm
 halo}<1.5\times 10^9h^{-1}\,M_\odot$) does not affect the value of $f_{\rm
 merg}$ for a cluster galaxy, but the effects of neglecting mergers with
sub-resolution 
haloes propagate to galaxies with masses up to nearly the critical mass,
$m_{\rm crit}=f_{\rm b}\,M_{\rm shock} = 1.3\times 10^{11}\,h^{-1}\,M_\odot$.

Reionisation and supernova feedback (the terms in $v_{\rm circ}$ in
Eq.~\ref{mstars}) produce an effect analogous to that of mass resolution by
suppressing galaxy formation in all haloes below $v_{\rm circ}=40\,\rm
km\,s^{-1}$ 
and by
considerably lowering the masses of galaxies in haloes just above this
threshold. Therefore, mergers with small haloes make no or a very small
contribution to the growth of the galaxy stellar mass,
which results in a strong suppression of $f_{\rm merg}$ at all masses
(\Fig{fmergers}, green dashed line vs. red dotted line). 

The suppression of star formation at halo masses greater than $M_{\rm shock}$
has a small effect on the merging histories of galaxies with $M_{\rm
 halo}\lsim M_{\rm shock}$ (blue dash-dotted line vs. red dotted line in
\Fig{fmergers}), but it 
means that mergers provide the only 
mechanism for galaxies to grow above the limit mass $m_{\rm stars}=m_{\rm
 crit}$. Therefore, including the cut-off at $M_{\rm shock}$ increases the
fractional importance of mergers in the growth of galaxies with $m_{\rm
 stars}>m_{\rm crit}$ (dash-dotted line relative to dotted line), 
not by affecting the merger rate
but by decreasing the importance of gas accretion.  Combining the terms in
$v_{\rm circ}$ and $M_{\rm shock}$ in equation~(\ref{mstars}) causes: i) a strong
suppression of the importance of mergers at $m_{\rm stars}<m_{\rm crit}$ and
ii) a strong increase in the importance of mergers at $m_{\rm stars}>m_{\rm
 crit}$, both with respect to a simple model where the baryons follow the
dark matter (dotted line).  This explains the
presence of two growth regimes, one dominated by gas accretion, the other
dominated by mergers, respectively below and above $m_{\rm stars}=m_{\rm crit}$.
Adding tidal stripping (solid line) has a minor effect on the median value
$f_{\rm merg}$, 
which may not be statistically significant.

\begin{figure}
% GAM SM: fmwetdryvsm fmcoldhot, fmwetdryvsm fmcoldhotE 0.4, fracE fracE
\centering
\includegraphics[angle=-90,width=0.85\hsize]{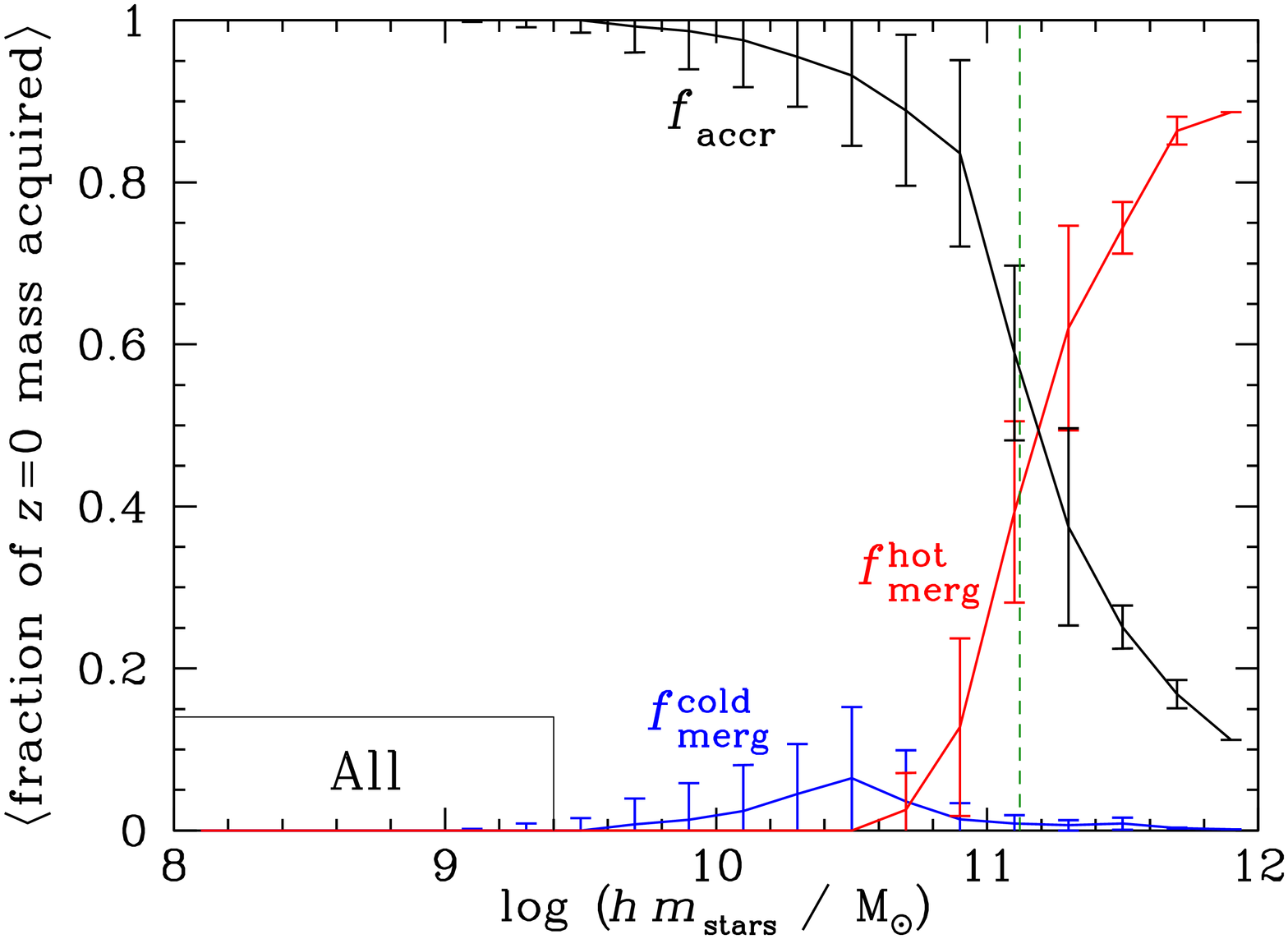}
\includegraphics[angle=-90,width=0.85\hsize]{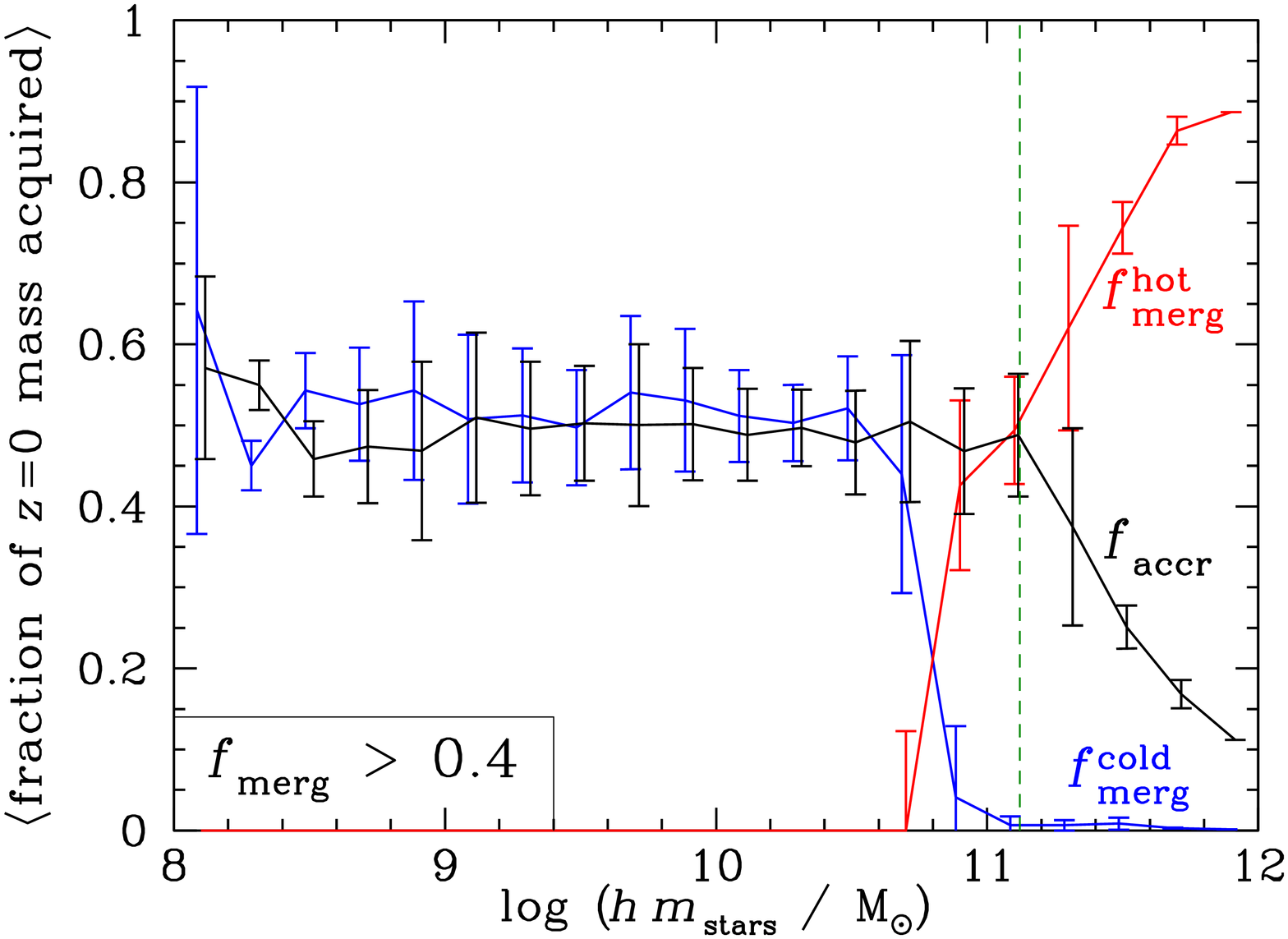}
\includegraphics[angle=-90,width=0.85\hsize]{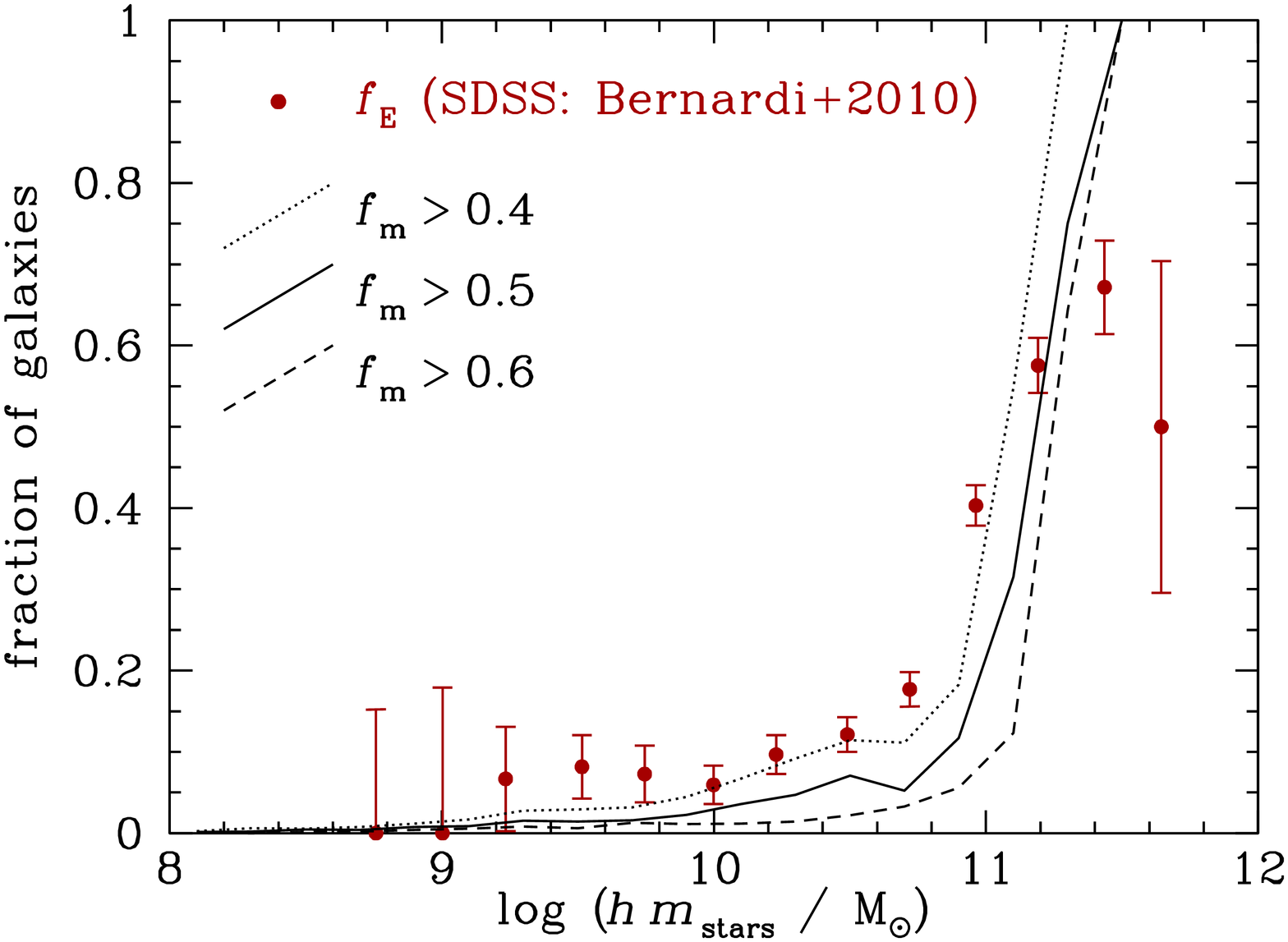}
\caption{
\emph{Top:} Median mass fraction at $z$=0 acquired via gas accretion (\emph{black line}), 
``cold mode'' galaxy mergers within haloes with $M_{\rm halo}^{\rm host}<M_{\rm shock}$ (\emph{blue line}),
or ``hot-mode'' ones within haloes with $M_{\rm halo}^{\rm host}>M_{\rm shock}$ (\emph{red line}).
Error bars show the interquartile range.
The \emph{green vertical dashed line} shows the baryonic mass $m_{\rm crit} = f_b M_{\rm shock}$.
\emph{Middle:} Same as top panel, but for galaxies with $f_{\rm merg} \geq
0.4$ (comparable to ellipticals, see next panel).
\emph{Bottom:} Fraction of galaxies with $f_{\rm merg} \geq f_{\rm m}$ for
$f_{\rm m} = 0.4$, 0.5, and 0.6.
The \emph{dark red circles} show the fraction of ellipticals, as visually
classified by \cite{fukugita_etal07} from SDSS images, as a function of
stellar mass, as presented by
\citeauthor{Bernardi_etal10} (\citeyear{Bernardi_etal10}, middle right panel
of their Fig.~12).
}
\label{fig:coldhotgas}
\end{figure}

\Fig{mstarvsMhalo} shows that $f_{\rm merg}$ increases with $m_{\rm stars}$
along the relation for central galaxies.  
The top panel of \Fig{coldhotgas} splits the mass
accretion history of galaxies into three channels: 1) gas accretion, 2)
\emph{cold-mode} mergers, and 3) \emph{hot-mode} mergers (see Section~\ref{Central and satellite galaxies}
for a precise definition of cold/hot-mode mergers).
In turns out that half of the 
galaxies with $m_{\rm stars}>10^{11}h^{-1}M_\odot$ have acquired at least half
of their mass via merging, and this merging involves almost always a primary
halo more massive than $M_{\rm shock}$.  On the other hand, most galaxies
with $m_{\rm stars} < 10^{11}h^{-1} M_\odot$ have acquired most of their mass
through gas accretion.  
This figure is virtually unchanged when we restrict ourselves
to the central galaxies.

The top panel of 
\Fig{coldhotgas} shows that
cold-mode mergers appear to contribute little to the
growth of galaxies.
In no mass bin is their median contribution to the final
stellar mass more than $\sim 5\%$.  This minor channel of galaxy mass growth
reaches its maximum relative significance at a galaxy stellar mass of order
$m_{\rm stars}=f_{\rm b}\,M_{\rm shock}$.  
In particular, among the satellites in groups and clusters with
$M_{\rm halo}^{\rm host} > 10^{13} h^{-1} M_\odot$, only 2\% have acquired
  over half their mass by mergers (6\% of the more massive ones with $m_{\rm
    stars} > 10^{10} \,h^{-1} M_\odot$ and 1\% of the ones with $10^9 M_\odot
  < h\,m_{\rm stars} < 10^{10} M_\odot$).

\begin{figure}
% GAM SM: fmergdistsnew fmergdists2 0.04
\centering
\includegraphics[width=0.9\hsize]{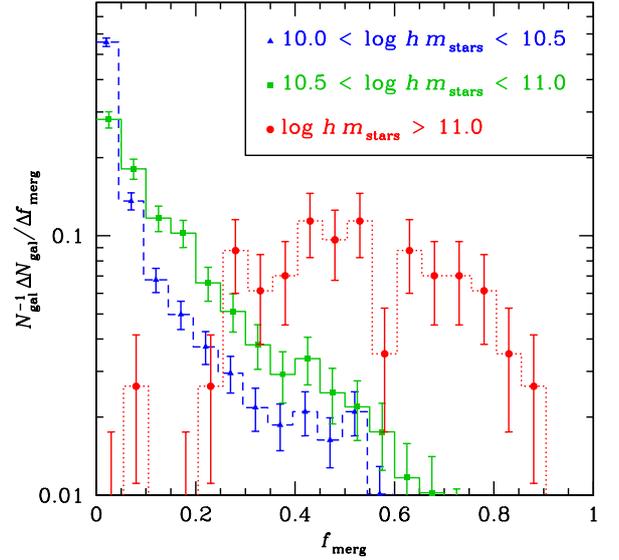}
\caption{Distribution of fraction of the  $z$=0 mass acquired by mergers for
  galaxies in 
 different bins of stellar mass.}
\label{fig:4histo}
\end{figure}

Still, some low-mass galaxies do acquire most of their their stellar mass via
mergers. Figure~\ref{fig:mstarvsMhalo} shows nearly one hundred central
galaxies with host halo masses below $M_ {\rm shock}$ that have acquired most
of their stellar mass by mergers, more precisely by cold-mode
mergers.\footnote{The large symbols for galaxies with high merger fractions in
  Figure~\ref{fig:mstarvsMhalo} creates an illusion of the dominance of
  cold-mode mergers in low mass ($10 < \log h\,m_{\rm stars}< 10.5$) central
  galaxies, which is dispelled in Figures~\ref{fig:coldhotgas} (top panel) and
  \ref{fig:4histo}.}   
\Fig{4histo} shows the distribution of the fractional contribution of mergers
to the final stellar mass for galaxies in four stellar mass intervals. 
A few galaxies that have accreted a significant fraction of their mass via
mergers are present even in the lowest mass bin.  In other words, by assuming
that a large mass fraction acquired by mergers implies elliptical galaxy
morphology (either through single major mergers or through repeated minor
mergers, as suggested by \citealp{bournaud_etal07}), our model
can accommodate the presence of $10^{10.5}M_\odot$ galaxies with early type
morphologies. 
 However, the probability that a galaxy has an early type
morphology (high $f_{\rm merg}$) is much higher at $m_{\rm
  stars}>10^{11}\,h^{-1}\,M_\odot$ than it is at $10^{10}M_\odot<h\,m_{\rm
  stars}<10^{10.5}M_\odot$.
As seen in the middle panel of \Fig{coldhotgas}, cold-mode mergers
thus do play an important role in the formation of those intermediate-mass galaxies
that mainly grow by mergers, and which may be compared to intermediate-mass ellipticals.

Finally, the bottom panel of \Fig{coldhotgas} shows that the fraction of
galaxies where the mass fraction acquired via mergers is $f_{\rm merg}\gsim 0.4$
matches reasonably well the
observed fraction of ellipticals for galaxy masses above $10^{10} h^{-1}M_\odot$
\footnote{The
  drop in the observed fraction of ellipticals at very high mass appears is
  caused by an increase of S0s, but the distinction between ellipticals and
  lenticulars is difficult and subject to errors.}.
 At $10^9 h^{-1}M_\odot<m_{\rm stars}<10^{10} h^{-1}M_\odot$, our model seems
 to predict less ellipticals than observed in the SDSS.
Our interpretation (based also on the structural properties of dwarf ellipticals and dwarf spheroidals) is that these objects were not formed
by mergers, but this occurs below our estimated mass
resolution limit. We defer further discussion of this point to the conclusion.

\begin{figure}
% GAM SM: fmergdistsnew fmergdists2 0.04
\centering
\includegraphics[angle=-90,width=\hsize]{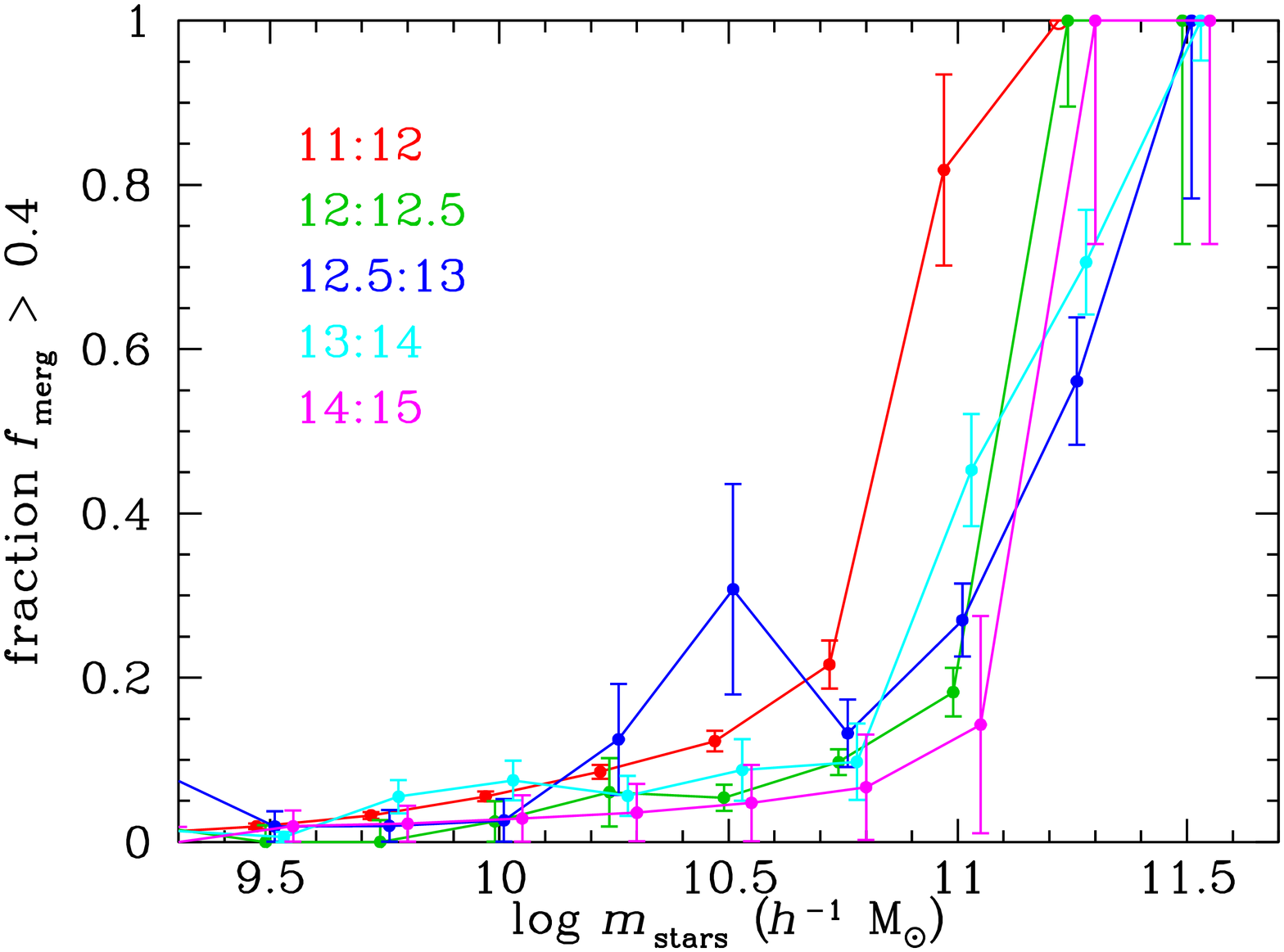}
\includegraphics[angle=-90,width=\hsize]{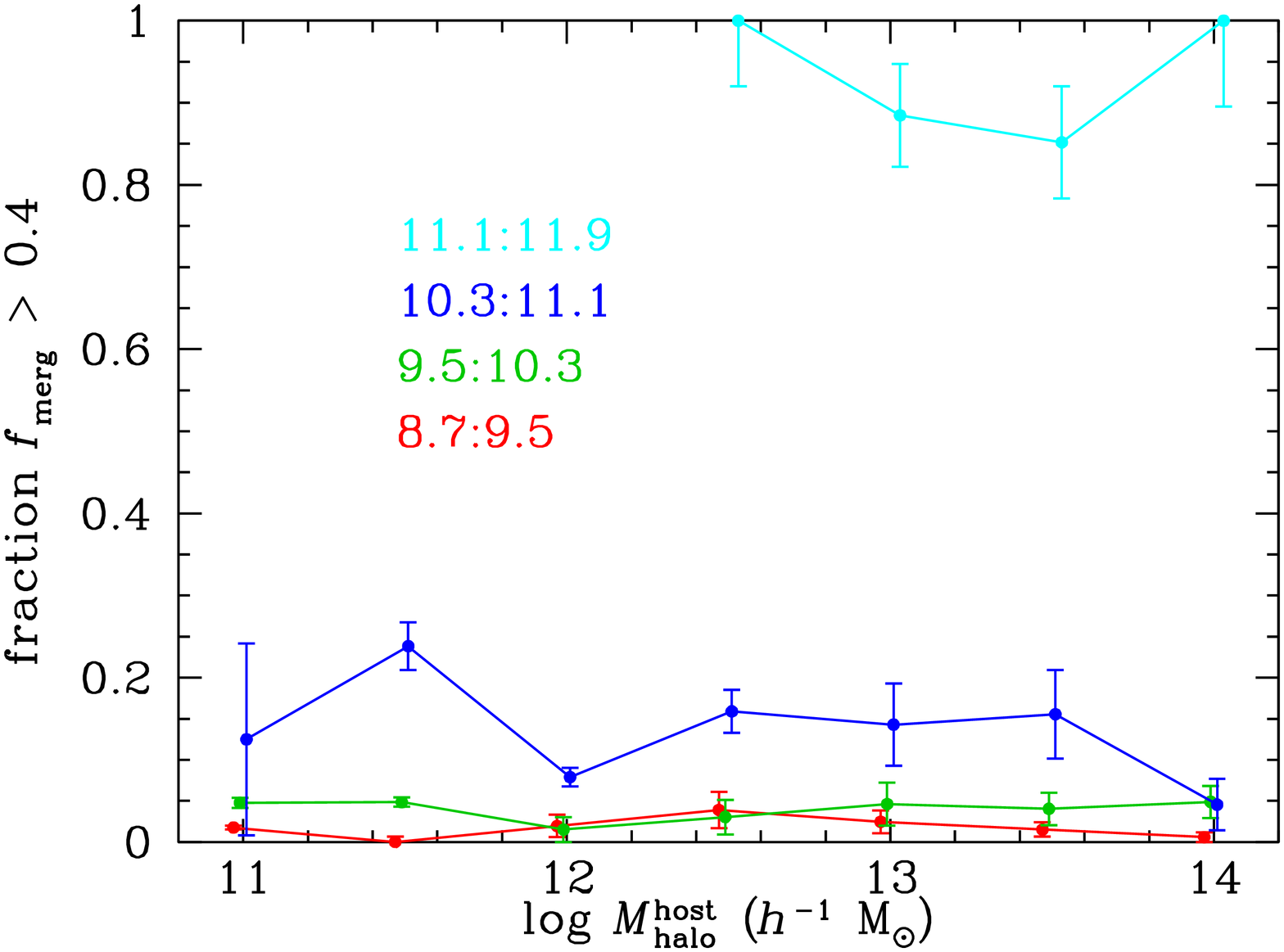}
\caption{\emph{Top}: 
Fraction of galaxies with over 40\% of their  $z$=0 mass acquired by
  mergers versus stellar mass in bins of halo mass (labelled by $\log h M$).
The errors are binomial.
\emph{Bottom}: Same versus halo mass in bins of stellar mass (labelled by
$\log h m_{\rm stars}$).
\label{fig:fmergvsmstars}}
\end{figure}

The top panel of \Fig{fmergvsmstars} shows the fraction of galaxies with over
40\% of the mass acquired by mergers (as the dotted line of the bottom panel
of \Fig{coldhotgas}), split in bins of halo mass. It is clear the halo
mass plays only a minor role in determining the fraction of galaxies with at least 40\% of
their mass acquired by mergers. In contrast, the bottom panel
of \Fig{fmergvsmstars} shows how the fraction of galaxies with over
40\% of the mass acquired by mergers varies with halo mass in bins of stellar
mass. 
If $f_{\rm merg} > 0.4$ is a proxy for elliptical galaxies (as can be 
inferred from
the bottom panel of \Fig{coldhotgas}), 
we predict that the fraction of ellipticals 
should depend more on the stellar mass of a galaxy than on its global
environment.

\begin{figure}
\centering
\includegraphics[width=\hsize]{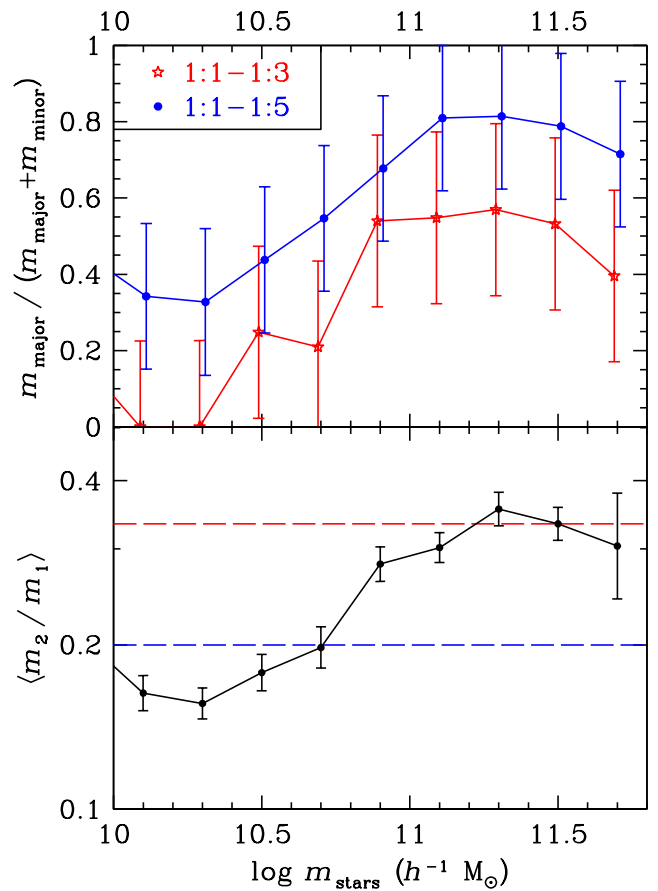}
\caption{\emph{Top}: 
median fraction of the mass at $z$=0 acquired by mergers due to major mergers for galaxies with $f_{\rm merg} > 0.4$.
The \emph{red} and \emph{blue lines} use stellar mass ratios between 1:1 and
respectively 1:3 and 1:5 to 
define major mergers. 
The \emph{error bars} are bootstrap estimates of the uncertainties of the mean fractions.
\emph{Bottom}: geometric mean mass ratio of mergers for galaxies with 
$f_{\rm merg}>0.4$ at $z=0$. 
The \emph{error bars} are bootstrap estimates of the uncertainties of the
geometric mean ratios.
The \emph{red} and \emph{blue horizontal lines} respectively represent the 1:3 and 1:5
limits of major mergers.
} 
\label{fig:majorminor}
\end{figure}

The top panel of \Fig{majorminor} shows the contribution of major mergers to the total mass
fraction at $z$=0 acquired via mergers for galaxies with $f_{\rm merg} >
0.4$ (hereafter, \emph{candidate elliptical} galaxies, following the bottom panel of
Fig.~\ref{fig:coldhotgas}).
Defining \emph{major} mergers as those
with stellar mass ratios between 1:1 and 1:3 (red lines in
\Fig{majorminor}), 
we find that 
at $m_{\rm
  stars} <  10^{10.8}h^{-1}M_\odot$ major mergers account for little of
the mass growth of candidate
ellipticals occurring through mergers.
But in galaxies with $m_{\rm stars}\gsim
10^{11}h^{-1}M_\odot$, major mergers contribute half of the mass growth by
mergers of candidate ellipticals.
If we
relax our definition of major mergers to include mergers with stellar mass
ratios between 1:1 and 1:5 (blue lines in \Fig{majorminor}), 80\% of the
merger mass growth of massive candidate
ellipticals ($m > 10^{11} h^{-1} M_\odot$) occurs through major
mergers, while at lower masses ($10 < \log h\,m < 11$), major mergers account
for at least one-third of the mass growth via mergers.

The bottom panel of \Fig{majorminor} shows the geometric mean of the 
mass ratios of the merging galaxies involved in the mass growth of candidate
ellipticals. Again one sees that the merger mass growth of candidate ellipticals is
mainly through minor mergers at low mass ($<10^{10.5} h^{-1} M_\odot$) and
major mergers at high mass ($>10^{11} h^{-1} M_\odot$). The importance of
major mergers peaks around $m_{\rm stars} = 10^{11.3} h^{-1} M_\odot$.

\section{Discussion and conclusion}
\label{discus}

Our goal is to understand how galaxies grow in mass, i.e. what are the
respective roles of gas accretion and galaxy mergers, as well as that of
feedback mechanisms in the
growth of galaxies.  

We have developed a \emph{toy model} of galaxy formation, using a very simple
hybrid
SAM/HOD approach to parameterise, as a function of halo mass and redshift, the
stellar mass present in galaxies after gas accretion and quenching of gas
infall by virial shocks and AGN and of star formation by reionisation and
supernovae.  We also include galaxy mergers, as measured from our
high-resolution cosmological N-body simulation. The stellar mass acquired via
gas accretion is modelled with a simple equation involving only three
parameters, while we added a fourth parameter to describe the stellar mass
loss caused by tidal stripping. We fined-tuned our four parameters to the
observed $z$=0 stellar mass function of galaxies.

In contrast with SAMs, our toy model does not incorporate structural or
morphological properties of galaxies, has no time derivatives built-in (for
the transfer of mass between galaxy components), and does not incorporate
luminosity, hence our galaxies have no colours.

Our choice to model the stellar mass as a unique function of halo mass and
redshift may seem oversimplified in view of the complexity of galaxy
formation, as exemplified by the very complex semi-analytical codes of galaxy
formation. However, at $\log hM_{\rm halo} > 12.15$ (where it
can be measured) the SDSS observations also appear to show a very small
scatter of $m_{\rm stars}$ vs. $M_{\rm halo}$ for central galaxies
(\citealp{yang_etal09} and the cyan region of our Fig.~\ref{fig:mstarvsMhalo}).
Moreover, both semi-analytical and hydrodynamical models of galaxy
formation indicate that central galaxies in halos show a very small scatter
between the stellar and halo masses (see Fig.~2 of
\citealp{cattaneo_etal07}), at $z=0$ and at $z=3$. 

The main worry with our
model is that our prescription for $m_{\rm stars}$ of equation~(\ref{mstars}) may
not be correct at $z>0$. However, our $m_{\rm stars} - M_{\rm
  halo}$ relation at $z=5$ (Fig.~\ref{sfevsM})  is not dramatically different
from that of \cite{behroozi_etal10}.
Moreover, we have reasons to believe that the cutoffs at
low and high mass should be of the right order, so the worry is on the
maximum efficiency (e.g. the maximum height of the top [red] curve of
Fig.~\ref{sfevsM}). Furthermore, our approximate match of the evolution of the
cosmic star formation rate density (Fig.~\ref{fig:cosmicSFR}) suggests that,
at $z>0$,
our model of equation~(\ref{mstars}) is of the right order
of magnitude. 

While we do not claim to have explored all possible parameter combinations,
any model that suppresses star formation even further will worsen our
agreement with the normalisation of the cosmic SFR density
(Figure~\ref{fig:cosmicSFR}).  We also note that the tendency to be at the
lower limit of the observational range for the cosmic SFR density is not
specific to our model, since it is also found in state-of-the-art
semianalytic models that are based on the same fundamental assumptions (see
Figure 8 in \citealp{cattaneo_etal06}).

\subsection{Tidal stripping}
\label{sec:discuss-tidal}

Without tidal stripping, we find that, even after correcting for
overmerging, we still have an excess of massive galaxies compared to the
\citet{bell_etal03} mass function (dotted line in \Fig{MFs}a).  This remains
true for other observed determinations of the galaxy stellar mass function
\citep{baldry_etal08,yang_etal09}.   
One could
argue that, although the semianalytic model that we use for the merger time
is a good match to what is seen in hydrodynamic simulations
\citep{jiang_etal08}, we may still overestimate the galaxy merger rate, which
is still somewhat uncertain.  However, it is difficult to lower the merger
rate without overproducing low-mass galaxies (see the dashed line in
\Fig{MFs}a).  One could imagine compensating for this effect by increasing
feedback at low masses, but we could not find a plausible parameter set that
does this and fits the galaxy mass function as well as the full model in
\Fig{MFs} does.

We believe that tidal stripping and observational errors in the SDSS
magnitudes that anchor the high end of the \citet{bell_etal03} mass function
(which according to \citealp{lauer_etal07} can run up to a magnitude) can
easily reconcile our model with the high end of the galaxy stellar mass
function.  The fundamental reason why the magnitudes of bright ellipticals
are difficult to measure is that their shallow light profiles do not
converge.  So it is hard to estimate the total light, let alone distinguish
galaxy light from intracluster light (see the discussion in the Appendix~A of
\citealp{cattaneo_etal08}).  Since the extended envelopes of massive
ellipticals are probably the products of interactions, the distinction
between the full model and the full model without stripping may be one more
of name than of substance.  If we decide that the envelopes belong to the
galaxies, then we should conclude that the full model without stripping is
correct and that the problem lies with the data, which are likely to
underestimate the real masses. If instead we decide that the envelopes are
intracluster light, then we should conclude that the model that includes
tidal stripping is the most physical one.  

Our best-fit stripped fraction per orbit, $\eta_{\rm
  strip}=0.4$, matches well the value determined in a recent dissipationless
simulation of a dwarf irregular galaxy orbiting the Milky Way with a fixed
potential 
\citep{Klimentowski+09}.\footnote{In a sequel with a live Milky Way potential,
\cite{lokas_etal10} find an even stronger decrease in stellar mass,  that
amounts to 0.65 per orbit among the particles within 1 kpc. Note, however,
that
the secondary
galaxy is 4 times more massive and with half the apocenter as the
respective values in the simulations of \cite{Klimentowski+09}.}
For $\eta_{\rm strip}=0.4$, the stripped mass in haloes with $M_{\rm halo}\sim
10^{13}-10^{14}h^{-1}M_\odot$ is in the range of $\sim 0.05-0.1 f_{\rm
  b}M_{\rm halo}$. This value is consistent with a number of independent
measurements: \citet{CastroRodriguez+09} find 7\% of diffuse intracluster
light (ICL) in the Virgo cluster, while \cite{FCJD04} find 15\% in Virgo; but
\cite{LinMohr04} find as much as 50\% in clusters, suggesting that $\eta_{\rm
  strip}$ might be even higher than we found.

\subsection{The role of feedback}

In agreement with 
previous studies, we find that in all haloes, but more so in low ($M_{\rm
  halo}\ll M_{\rm shock}$) and high ($M_{\rm halo}> M_{\rm shock}$) mass
haloes, $m_{\rm stars}/M_{\rm halo}$ must be much smaller than the universal
cosmic baryon fraction if the halo mass function predicted by the CDM model
is to be reconciled with the observed galaxy mass function.

The mechanism that suppresses star formation in haloes with $M_{\rm halo}>
M_{\rm shock}$ is clear. 
The post-shock gas is hot enough to
maintain a stable shock, while in lower mass halos, the post-shock gas
radiates efficiently and can no longer provide the thermal pressure to
support the shock and collapses instead.
The question is why can't the  gas around massive haloes cool
down again.  Growing evidence suggests that this is due to heating from the
central AGN (see \citealp{cattaneo_etal09}; and references therein).

Reionisation and supernova feedback are widely invoked to explain the very small
$m_{\rm stars}/M_{\rm halo}$ ratios of galaxies in low mass haloes (e.g.,
\citealp{mcgaugh_etal10}). The trouble is that 
attempts to simulate these processes have difficulty to produce outflows as
large as those that are required by this study because the energy that is
available is used inefficiently \citep{DT08,CK09}. However, unless the CDM model is
wrong, whatever other process may be relevant must behave analogously to the
reionisation and supernova feedback model described in the article, if it is
to fit the observed stellar mass function \citep{bell_etal03} that we have used.

The fact that $m_{\rm stars}/M_{\rm halo}$ drops at halo masses below and
above $M_{\rm shock}$ introduces a characteristic scale for galaxy masses
that makes them deviate from the dark matter's self-similar evolution
\citep{marinoni_hudson02,baldry_etal08,li_white09}.  If the evolution of
baryons followed that of the dark matter, then galaxies of all masses would
have similar merging histories (to the extent that dark matter halos evolve
self-similarly, which is true to first order, e.g. \citealp{vandenBosch02}).  
Instead, it is clear that the baryonic physics, whatever
they are, break the dark matter's scale-invariant behaviour by suppressing
star formation in low and high-mass haloes.  It is only within haloes in a
narrow mass-range around $M_{\rm halo}\sim M_{\rm shock}\sim
10^{12}h^{-1}M_\odot$ that baryons can form stars efficiently (the galaxy
formation zone in Fig.~\ref{fig:sfe}).

This conclusion is similar to that of \cite{bouche_etal10}, who developed a
toy model similar to ours in that it also incorporates minimum and
maximum halo masses for efficient galaxy formation, although their emphasis
is different. They argue that a fixed halo mass floor of $10^{11} M_\odot$
for galaxy formation appears to be required to
reproduce the observations of specific star 
formation rate as a function of stellar mass, the evolution of the cosmic
star formation rate
and the
Tully-Fisher relation.
Our model is more focussed on the accurate treatment of the role of galaxy
mergers in the growth of galaxies, than on the modeling of star formation
rates. However, our model also
allows for lower
galaxy masses at high redshift, which may help 
explain the lowest mass dwarf galaxies.

\subsection{Mergers vs. gas accretion}

The shutdown of gas accretion above the critical mass $M_{\rm
  shock}\sim 10^{12}M_\odot$ is the reason why, in our model,
  mergers are the only opportunity
for galaxies to grow above the limiting mass $m_{\rm crit}\equiv f_{\rm
  b}\,M_{\rm shock}$. Therefore, galaxies with $m_{\rm gal}\gg m_{\rm crit}$
must have acquired a significant fraction of their mass via mergers.  
This conclusion appears quite robust because without this shutdown 
massive galaxies would be too frequent and too blue
\citep{bower_etal06,cattaneo_etal06,croton_etal06}.

Where does the dearth of mergers at galaxy masses lower than $10^{11} h^{-1}M_\odot$ come from?
It cannot be a resolution effect (though finite resolution makes it more pronounced)
because even at $m_{\rm stars}>10^{10.6} h^{-1} M_\odot$ the mass fraction acquired via mergers rises
faster with galaxy mass in our best-fit model than when we force the baryons to
trace the dark matter.
There is, instead, a simple explanation why the importance of mergers decreases at low masses.
The steeper slope of stellar
versus halo mass at low mass (Fig.~\ref{fig:mstarvsMhalo})
naturally reduces the relative abundance of major
partners relative to minor ones, and therefore the mass contribution that any merger can make.
\cite{hopkins_etal10} have made a similar kind of argument.

The present work reinforces the new picture in which most galaxies grow through
gas accretion and form stars along the \emph{Blue Cloud} until the critical
halo mass for shock and AGN heating is reached. Then, gas accretion is shut
off, star formation is quenched and they migrate to the \emph{Red Sequence},
where gas-poor mergers represent their only opportunity for growth
\citep{bower_etal06,cattaneo_etal06,croton_etal06,faber_etal07,cattaneo_etal08}.

We refer to mergers that take place in host haloes above $M_{\rm shock}$ as
\emph{hot-mode} mergers, and to mergers that take place in host haloes below
$M_{\rm shock}$ as \emph{cold-mode} mergers.  The former usually contain
little cold gas, while the latter involve galaxies containing large masses of
cold gas supplied by cold accretion.  It is thus tempting to identify
hot-mode and cold-mode mergers with gas-poor (dissipationless) and gas-rich
mergers, respectively.  The latter are mainly confined to galaxies below the
critical stellar mass $f_{\rm b}\,M_{\rm shock}$
(Fig.~\ref{fig:mstarvsMhalo}) and are most frequent just below $f_{\rm
  b}\,M_{\rm shock}$ (top panel of Fig.~\ref{fig:coldhotgas}).  This points to a
characteristic mass of $\sim 10^{10.5}h^{-1}\,M_\odot$ for ultraluminous
infrared galaxies (ULIRGs, e.g. \citealp{colina_etal01}) and quasar hosts
(e.g. \citealp{bonoli_etal09}), of which gas-rich mergers are the supposed
triggering mechanism (top panel of \Fig{coldhotgas}).

Our model, therefore, predicts three regimes of galaxy formation (see top
panel of \Fig{coldhotgas}):
Galaxies with $m_{\rm stars}\lsim 10^{10}\,M_\odot$ were effectively built
up by gas accretion only.
At $10^{10}\,M_\odot\lsim m_{\rm stars}\lsim 10^{11}\,M_\odot$ gas accretion
remains the dominant mechanism, but we also see a population that was built by
gas-rich mergers.
Galaxies with $m_{\rm stars}\gsim 10^{11}M_\odot$ were mainly built by gas-poor
mergers,
in agreement with the analysis of SDSS galaxies by
\cite{BRSS11}, who explain the high-mass trends of color, elongation and color
gradient of SDSS ellipticals with gas-poor mergers, by invoking a transition at the same
stellar mass.
Galaxies built by gas-rich mergers are rare at all masses. However, as expected
from our model (middle panel of Fig.~\ref{fig:coldhotgas}), cold-mode
mergers dominate the growth of those galaxies with $m_{\rm stars}<
10^{11}\odot$ that are mainly built
by mergers.

Running a semi-analytical model on top of the Millennium Simulation,
\cite{delucia_etal06} find that the `effective' number of progenitors (a
proxy for the number of major progenitors) of galaxies
increases sharply beyond $10^{11} M_\odot$, which appears to be consistent with our conclusion
that high-mass galaxies are built by major mergers.

Analyzing the same Millennium Simulation (which
has 11 times worse mass resolution than our simulation), \cite{guo_white08}
measured the rate of growth of galaxies in the channels of gas accretion
(which they denote ``star formation''), and major and minor mergers. They
also found that while major mergers dominate the growth of the most massive
present-day galaxies, mergers contribute negligibly to the growth of low
present-mass galaxies. Moreover, they noticed, as we did (our
Fig.~\ref{fig:fmergers}) that the importance of mergers increases with final
galaxy mass in a stronger way than do mergers of the dark matter haloes.

\cite{hopkins_etal10} followed an alternative approach
to analyse the role of
mergers in the growth of galaxies, in particular the relative importance of
major and minor mergers.
They used the halo merger rate function that \cite{fakhouri_ma08}
derived from N-body simulations together with abundance matching to
associate stellar masses to the merging haloes.
The delay between halo mergers and galaxy mergers was calculated with
the fitting formula by \cite{boylankolchin_etal08},  which is
analogous to the \cite{jiang_etal08} formula that we use here.
Their results for the relative importance of major mergers are, to
first order,
similar to ours: 1) major mergers dominate the formation and assembly of
$m_{\rm stars}\sim m_{\rm crit}$ ellipticals, but the contribution of  minor
mergers is non-negligible $\sim 30\%$; 2) the formation of
elliptical galaxies becomes dominated by minor mergers at lower
masses;
But \citeauthor{hopkins_etal10} find that  the relative
importance of major mergers rapidly decreases again at high 
galaxy masses $m_{\rm stars} > 10^{11} M_\odot$,
while we find a small and not statistically significant drop in the
importance of major mergers at very high galaxy masses ($\log h\,m_{\rm
  stars} > 11.5$). 
When we consider
the geometric mean mass ratio as a function of stellar mass, which is more
comparable to the median mass ratio of bulges considered by
\citeauthor{hopkins_etal10},
we obtain
(bottom panel of \Fig{majorminor}) a globally similar trend as
\citeauthor{hopkins_etal10}, although again with two differences:
1) their peak for for major mergers occurs at 3 times lower galaxy mass (after
correcting for $h$); 2) the drop in merger mass ratio is much more pronounced
in their model than in ours.

It is difficult to say which result is more correct in quantitative
detail.
The approach followed by \citeauthor{hopkins_etal10} is certainly more
accurate in its capacity to predict the galaxy merger rate at a given
time,
since the halo occupation distribution is directly derived from
abundance matching.
However, 
our approach follows more consistently the
merger history across the Hubble time of each halo inside the N-body simulation.

\citet{maller_etal06}
analysed the problem directly with hydrodynamic cosmological simulations.
They
concluded that mergers are more important at high galaxy masses.
However, their simulations lack effective feedback and substantially
overestimate the observed galaxy mass function.  Therefore, they
overestimated the role of gas accretion.  Nevertheless, they determined
that the average number of both minor and major mergers increases with
$m_{\rm stars}$.  They found that their high mass galaxies have typically undergone
one major (1:1 to 1:4) merger through their lifetime. We note that the
 threshold they use to define high-mass galaxies is low ($\log h\,m_{\rm stars}
 = 10.6$), so we expect
that the dominance of mergers in their high mass bin is diluted by the importance of
accretion near this mass threshold. Not only do
we find that mergers are common among the most massive galaxies.  We also
find that the masses of these galaxies were mainly built by mergers.

\citet{naab_etal07} also used hydrodynamic simulations to address the
importance of mergers.  However, in contrast with \citet{maller_etal06}, who
simulated a cosmic volume at low resolution, they used very high resolution
to zoom into the formation of three individual ellipticals.  They found that their
galaxies have grown by mergers from $z$=1 to $z$=0 by $25\%$ in mass on
average.  They, therefore, concluded that intermediate-mass elliptical
galaxies were not built by major mergers.  We note, however, that their three
ellipticals accreted over half of their mass by mergers since $z$=5.
Moreover, at $z$=0, these galaxies live in haloes with virial masses in the
range of $1.6-2.3\times 10^{12}\,M_\odot$.  For central galaxies, this halo
mass range corresponds to $m_{\rm stars} \simeq 10^{11.0} h^{-1} M_\odot$,
which is the precise galaxy mass for the transition in mass acquisition 1) by gas
accretion and by mergers (\Fig{coldhotgas}), and 2) by minor and major
mergers 
(Fig.~\ref{fig:majorminor}).  We therefore find the agreement between our
results and those by \citet{naab_etal07} quite encouraging.

The increasing importance of mergers on one hand and the importance of 
major mergers on the
other are confirmed through two  analytical developments that are 
deferred to a forthcoming paper.

Finally, we remark that the delay of galaxy mergers with respect to
halo mergers, which we compute with the the dynamical friction formula
in the  \cite{jiang_etal08} version,
turns out to play a very modest role in our conclusions:
the galaxies that have not yet
merged with the central galaxies of their direct hosts, even
though their galactic haloes have merged with their immediate host
haloes (or have become unresolved because of tidal stripping of most of their
mass), account for $\lsim 1\%$ of the total
galaxy population and $\sim 10\%$ of the satellite galaxy population
at $m_{\rm stars}>10^9{\rm\,h}^{-1}M_\odot$ (satellites
make up $\sim 10-15\%$ of all galaxies).

\subsection{The dichotomy between central and satellite galaxies}

Our very simple toy model of galaxy formation shows very distinct properties
for central and satellite galaxies (see Fig.~\ref{fig:mstarvsMhalo}). Central galaxy mass increases slightly with halo mass, while the galaxy mass
function, when measured within a narrow range of halo mass, clearly
separates the central population and the satellite population, with a sharp peak at high-masses that corresponds to
central galaxies (top panel of Fig.~\ref{masfuns}). This feature was first
predicted by \cite{Benson+03} and \cite{zheng_etal05} and first clearly 
observed by
\cite{yang_etal09} in the SDSS survey, who also modelled it with
the conditional stellar mass function formalism. As we wrote these
lines, we learnt that \cite{liu_etal10} were also able to reproduce this
behaviour with a semi-analytical model.

Over three-quarters\footnote{The 22\% of non-Red Sequence cluster galaxies
  found by \cite{yang_etal08} matches perfectly the fraction of interlopers
  projected along the virial sphere, i.e. the average fraction of particles
  in the virial cone that are outside the virial sphere
  \citep{MBM10}. Hence, the fraction of non-Red Sequence galaxies
  within the virial sphere is probably much lower than one-quarter. Indeed,
  the observed increase of the fraction of recent star forming galaxies
  (RSBGs) with projected radius has been deprojected by \cite{mahajan_etal10}
  to yield a fraction of $13\pm1\%$ of RSBGs within the virial spheres in
  comparison with $18\pm1\%$ within the virial cones.} of the galaxies in
clusters, the huge majority of which are satellites, are observed to lie on
the Red Sequence \citep{yang_etal08}, which is visible to very low
luminosities (e.g. \citealp{boue_etal08}) and is mainly composed of galaxies
with early-type morphologies, the great majority of which are classified as
dwarf ellipticals (dEs).  

Since mergers are found to be unimportant for the growth of most satellites
(Fig.~\ref{fig:mstarvsMhalo}), including those in groups and clusters (Sect.~\ref{sec:mergers}), one is led
to the conclusion that dEs must also acquire their mass by gas accretion.  Therefore,
the morphological properties and the shutdown of star formation in dEs (and
by extension in
dwarf spheroidals) must be due to processes unrelated to mergers and not included in our
model (which does not describe the conversion of gas into stars), such as ram-pressure
stripping \citep{GG72}, starvation \citep{LTC80} and harassment by repeated fast encounters
\citep{moore_etal98,mastropietro_etal05}.  

Therefore, we conclude that the mass-growth of elliptical
galaxies is a function of their present-day mass:
the most massive ones ($m_{\rm stars} > 10^{11} h^{-1} M_\odot$) have mainly
grown by gas-poor mergers, intermediate-mass ellipticals ($ 10^{10} h^{-1} M_\odot< m_{\rm stars} <
10^{11} h^{-1} M_\odot$) by gas-rich mergers, and the low-mass ones
($m_{\rm stars} < 10^{10} h^{-1} M_\odot$) by gas accretion and later
transformed by non-merging processes such as ram pressure stripping or galaxy
harassment. 
This is similar to the picture for all galaxies, except that
intermediate-mass ellipticals grow by gas-rich mergers while intermediate-mass
spirals are built by gas accretion.

\subsection{Caveats}

An important caveat to our analysis is that our model does not describe star
formation rates, as we are only interested in the total mass of a galaxy and the
fraction of this mass that comes from mergers.  This paper is not concerned
with the fraction of the mass that is in gas and the fraction of the mass
that is in stars. The SFRs that we show (Figure~\ref{fig:cosmicSFR}) are
simply mass accretion rates that are estimated assuming that all the accreted
gas is instantaneously turned into stars.  Therefore, it is not surprising that
our model anticipates the peak of the star formation rate in comparison with
observations (\Fig{cosmicSFR}).
In practice, at redshift $z\sim
0$, where we compare our model with the data (Figures~\ref{fig:MFs}a and
\ref{fig:mstarvsMhalo}), the gas is usually a small fraction of the galaxy
mass for intermediate and high-mass galaxies (see, e.g., 
\citealp{mcgaugh_etal10} and
references therein).  Therefore, the 
error that we make by identifying the stellar mass with the total galaxy mass
(stars plus cold gas) is at most $\sim 10-20\%$, which is compatible with the
level of accuracy that we expect from our model.  We note that even
observationally, the cosmic SFR density derived from dust-corrected measured
SFRs deviates at $z>1$ from the time derivative of the cosmic stellar mass
density (\citealp{wilkins_etal08}; Figure~\ref{fig:cosmicSFR}).  The origin
of this discrepancy is an open problem.  Our model matches the observed
time derivative of the comic stellar mass density much better than it matches
the measured star formation rate density.

Much higher mass resolution on our side would certainly be desirable to
properly follow substructures until they have merged with their hosts.
However, this problem has been partly handled with the incorporation of
delayed merging on a dynamical friction timescale
(Sect.~\ref{sec:galaxy_mergers}).  Moreover,
we are confident that our finite mass resolution is sufficient to indicate
the strong drop in the importance of mergers at masses below $10^{11} h^{-1}
M_\odot$ (Fig.~\ref{fig:fmergers}).  Also, our low $v_{\rm circ}$ cut-off
implies that there is not 
much advantage at resolving haloes with masses that are much lower than the
mass at which $v_{\rm circ}=v_{\rm reion}$, because there is no formation of
galaxies in those haloes after the epoch of reionisation.

Of course, the details of our toy galaxy formation model are most probably
oversimplified. The trend seen in the relation of galaxy mass to halo mass
(Fig.~\ref{fig:mstarvsMhalo}) might be incorrect at low
masses (although there is good agreement with the relation that
\citealp{GWLB10} derived
from abundance matching, see Fig.~\ref{fig:mstarvsMhalo}). 
Nevertheless, none of our conclusions seem to depend on the accuracy
of the model at low masses.

\subsection{Open questions}

Mergers are a reality and we have tried to present a careful and robust
estimate of their importance.  Our work suggests that present-day giant
ellipticals can only be built by gas-poor mergers, as gas accretion must be quenched at
high halo mass to avoid over-massive galaxies.  
Moreover, intermediate-mass ellipticals
have probably grown by gas-rich mergers, while mergers appear unimportant for
dwarf ellipticals.
These conclusions are consistent with the structural properties
of ellipticals: the surface brightness profiles of massive ellipticals tend
to display inner cores, while those of less massive ellipticals tend to be
cuspy \citep{kormendy_etal09}.

Still many questions remain.  
Can the well-defined relation between the masses of central supermassive
black holes and the 
spheroids of their hosts \citep{magorrian_etal98} be understood with both gas-poor
mergers at the high mass end and gas-rich mergers at intermediate masses?
Can this relation be extended from ellipticals to spiral galaxies? 
How do mergers fit in the growth of galaxies viewed at much higher redshift?
Are other mechanisms responsible for the growth of  
proto-ellipticals at high redshift
(e.g. \citealp{dekel_etal09})?

\begin{acknowledgements}

AC thanks the IAP for its hospitality during numerous visits.
AC and GAM acknowledge stimulating conversations with M. Bernardi,
D. Ceverino,
I. Chilingarian, A. Dekel, S.M. Faber, J.P. Ostriker, J. Silk, D. Tweed, F. van den Bosch,
and D. Weinberg.  We acknowledge anonymous referees for numerous
useful comments.  We
also thank C. Wagner for running our simulation, as well as A. Hopkins and
M. Bernardi for providing the observational data for
Figs.~\ref{fig:cosmicSFR} and \ref{fig:coldhotgas}, respectively.  AK is
supported by the MICINN through the Ramon y Cajal programme.  KW acknowledges
support through the DFG grant KN 755/1.

\end{acknowledgements}

\bibliography{ref_av}
\end{document}